%% file: main.tex
\documentclass[aps,prd,twocolumn,amsmath,amssymb,nofootinbib,superscriptaddress,floatfix,10pt]{revtex4-2}

\usepackage{graphicx}
\usepackage[colorlinks=true,linkcolor=blue]{hyperref}
\usepackage{tikz}
\usepackage{xcolor}
\usepackage[capitalise]{cleveref}

\newcommand{\B}{\mathcal{B}}
\newcommand{\C}{\mathbb{C}}
\newcommand{\CL}{\mathrm{CL}}
\newcommand{\eps}{\varepsilon}
\newcommand{\ii}{\mathrm{i}}
\newcommand{\im}{\mathrm{Im}}
\newcommand{\Nmeas}{N_{\mathrm{meas}}}
\newcommand{\Nruns}{N_{\mathrm{runs}}}
\newcommand{\Nsim}{N_{\mathrm{sim}}}
\newcommand{\obs}{\mathcal{O}}
\newcommand{\R}{\mathbb{R}}
\newcommand{\re}{\mathrm{Re}}
\newcommand{\taumax}{\tau_{\max}}
\newcommand{\taumeas}{\tau_{\mathrm{meas}}}
\newcommand{\tautherm}{\tau_{\mathrm{therm}}}

\Crefname{section}{Sec.}{Secs.}

\DeclareMathOperator{\sign}{sign}

\graphicspath{{./figures/}}

\begin{document}

    \title{The Role of Integration Cycles in Complex Langevin Simulations}

    \author{Michael W. Hansen}
    \email{michael.hansen@uni-graz.at}
    \affiliation{Institute of Physics, NAWI Graz, University of Graz, Universitätsplatz 5, 8010 Graz, Austria}
    \author{Michael Mandl}
    \email{michael.mandl@uni-graz.at}
    \affiliation{Institute of Physics, NAWI Graz, University of Graz, Universitätsplatz 5, 8010 Graz, Austria}
    \author{Erhard Seiler}
    \email{ehs@mpp.mpg.de}
    \affiliation{Max-Planck-Institut für Physik (Werner-Heisenberg-Institut), Boltzmannstraße 8, 85748 Garching bei München, Germany}
    \author{Dénes Sexty}
    \email{denes.sexty@uni-graz.at}
    \affiliation{Institute of Physics, NAWI Graz, University of Graz, Universitätsplatz 5, 8010 Graz, Austria}

\begin{abstract}	
	Complex Langevin simulations are an attempt to solve the sign (or complex-action) problem 
    encountered in various physical systems of interest. The method is based on a complexification 
    of the underlying degrees of freedom and an evolution in an auxiliary time dimension. The 
    complexification, however, does not come without drawbacks, the most severe of which is the 
    infamous `wrong convergence' problem, stating that complex Langevin simulations sometimes fail 
    to produce correct answers despite their apparent convergence. It has long been realized that 
    wrong convergence may -- in principle -- be fixed by the introduction of a suitable kernel 
    into the complex Langevin equation, such that the conventional correctness criteria are met. 
    However, as we discuss in this work, complex Langevin results may -- especially in the 
    presence of a kernel -- still be affected by unwanted so-called integration cycles of the theory 
    spoiling them.  Indeed, we confirm numerically that in the absence of boundary terms the complex 
    Langevin results are given by a linear combination of such integration cycles, as put forward by 
    Salcedo \& Seiler \cite{SS19}. In particular, we shed light on the way different choices of kernel 
    affect which integration cycles are being sampled in a simulation and how this knowledge can be 
    used to ensure correct convergence in simple toy models.
\end{abstract}	

\maketitle

\section{Introduction}\label{sec:introduction}
    \input{introduction}

\section{Complex Langevin evolution}\label{sec:cle}
    \input{cle}

\section{Integration cycles}\label{sec:cycles}
    \input{cycles}

\section{Simulation setup}\label{sec:simulation_setup}
    \input{simulation_setup}

\section{Results}\label{sec:results}
    \input{results}

\section{Summary and conclusions}
    \input{conclusions}

\begin{acknowledgments}
We express our gratitude towards Enno Carstensen and Ion-Olimpiu Stamatescu for valuable 
discussions on a regular basis and for fruitful collaborations past and present. The numerical 
results presented in this work have been obtained in part in simulations on the computing cluster 
of the University of Graz (GSC) and the Vienna Scientific Cluster (VSC). This research was funded 
in part by the Austrian Science Fund (FWF) via the Principal Investigator Project 
\href{https://doi.org/10.55776/P36875}{P36875}. The analyses performed in this work are based on 
the \verb|Python| ecosystem for scientific computing \cite{python}, explicitly through the use of 
the packages \cite{numpy,scipy,pandas,matplotlib,cmcrameri}, but we are also grateful for the 
creation and maintenance of all their dependencies. Moreover, the numerical integrations were 
computed with the help of the \verb|Wolfram Mathematica| software system \cite{mathematica}.
\end{acknowledgments}

\section*{Open Access Statement}
For the purpose of open access, the authors have applied a Creative Commons Attribution (CC BY) 
license to any author-accepted manuscript version arising.

\section*{Data Availability Statement}
The full data set underlying this work is available under \cite{data} and the analysis scripts
used are published online as well \cite{scripts}. Moreover, our simulation code is available upon
request.

\appendix
\section{Counting integration cycles}\label{app:counting_cycles}
    \input{counting_cycles.tex}
\section{Testing the goodness of fits}\label{app:chi_squared}
    \input{chi_squared.tex}

\bibliographystyle{apsrev4-2}
\bibliography{bibliography}

\end{document}

%% file: introduction.tex
Lattice field theory is the most successful method to date for studying 
nonperturbative aspects of quantum field theory from first principles. One of its
major achievements, for instance, is the discovery that the finite-temperature 
transition in Quantum Chromodynamics (QCD) is, in fact, an analytic crossover 
\cite{AEF06}. The fact that the temperature axis of the QCD phase diagram
is so well understood by now is owed to a large extent to lattice simulations by various
collaborations around the globe.

Conventional lattice field theory simulations are based on the idea of importance 
sampling, by which the Boltzmann factor $e^{-S_E}$ in the Euclidean path integral 
(with $S_E$ denoting the Euclidean action of the theory under study) is interpreted
as a probability density from which one draws samples of field configurations.
This approach, however, fails if $S_E$ has a nonvanishing imaginary part
since then this probabilistic interpretation breaks down. This is the case, for
instance, in QCD with a nonzero baryon chemical potential, which is the reason 
why straightforward lattice simulations using importance sampling cannot be 
applied to the study of QCD at finite baryon density \cite{TW05}. This 
so-called sign (or complex action) problem is not just restricted to 
finite-density QCD but is also present in various other systems, such as 
real-time quantum field theory or gauge theories with a $\theta$ term.

Needless to say, there are countless attempts for solving or at least mitigating 
the sign problem, which can be roughly classified into the following categories:
reweighting (e.g., classical reweighting \cite{FKT03} or density-of-states 
\cite{Gok88} methods), reformulation of theories in terms of new variables (e.g., 
the worldline approach \cite{Cha09p}), methods based
on complexification of the underlying field manifold (e.g., the complex Langevin
\cite{Par83,Kla83} or Lefschetz thimble \cite{CDS12} approaches) or approaches 
based on the Hamiltonian 
formalism (e.g., tensor networks \cite{BCC18}). Moreover, for QCD in the presence of 
a (baryon) chemical potential $\mu$, two of the most successful methods that are 
commonly employed are the analytic continuation from simulations at imaginary 
$\mu$, for which there is no sign problem \cite{AKW99}, and Taylor expansions in
$\mu/T$ (with $T$ denoting the temperature) around $\mu=0$ \cite{AEH02}. 

While all of these ideas come with their own advantages as well as drawbacks, in 
this work we shall focus on the complex Langevin approach. Based on extending the
idea of stochastic quantization \cite{PW81} (see also \cite{DH87r,Nam93p} for
reviews) to complex actions, this method found some success initially 
\cite{KW85,Sod88,GL93,OSZ93p} but was largely discarded afterwards as it caused 
problems that seemed insurmountable \cite{AY85,AFP86,Sal93,FOS94} at the time. 
However, the idea re-emerged in an attempt to enable simulations of real-time quantum 
field theories out of equilibrium \cite{BS05,BBS07} and has been developed further 
since. In particular, the addition of adaptive step-size algorithms \cite{AJS10} and 
gauge cooling \cite{SSS13} to the complex-Langevin toolbox enabled exploratory studies 
of finite-density QCD \cite{Sex14,Sex14_2,ASS14,FKS15,AJZ22u} as well as the 
determination of the phase diagram of heavy-dense QCD \cite{AAJ16}; see also 
\cite{BRL21} for a recent review.

Moreover, important progress has been made in assessing the convergence properties
of complex Langevin simulations \cite{ASS10,AJS11,Sal16,NNS16}. In particular, the
study of boundary terms has shed some light on the question why the complex Langevin
evolution sometimes produces incorrect results despite a reasonable convergence 
\cite{SSS19,SSS20}. However, it has also become clear that boundary terms on
their own are in general not sufficient as a correctness criterion \cite{SSS24}. 
Indeed, it was shown in \cite{SS19} that (at least in one dimension) the absence of 
boundary terms does not guarantee correct convergence, but only implies that the 
results obtained in a simulation are linear combinations involving different so-called 
integration cycles (using the terminology of \cite{Wit11,Wit10}). 

In this work, we shall elaborate more on the role played by integration cycles in 
the context of complex Langevin simulations and, in particular, how the introduction 
of a kernel into the Langevin equation can affect this picture. In \cref{sec:cle} we 
introduce the (complex) Langevin equation as well as the concept of boundary terms. 
We then devote \cref{sec:cycles} to a thorough discussion of integration cycles before
outlining our simulation setup in \cref{sec:simulation_setup}. There, we also briefly 
describe our efforts regarding the reproducibility of our results along the FAIR guiding 
principles \cite{FAIR16}. Lastly, in \cref{sec:results}, we present simulation results 
for simple one- and two-dimensional toy models to elucidate the relation between the 
kernel and the cycles that are being sampled in simulations. A number of auxiliary 
results are provided in the appendices.

%% file: cle.tex
As was mentioned in the Introduction, the sign problem in lattice quantum field theory 
arises when the weight $\rho[x]:=e^{-S[x]}/Z$ (here and in the remainder of this work we 
shall drop the subscript `$E$') in the path integral is not real and non-negative. Here, 
the partition function $Z$ is defined as
\begin{equation}
    Z := \int_M \mathcal{D}x\,e^{-S[x]}\;,
\end{equation}
where $x$ collectively denotes all dynamical degrees of freedom of the theory of interest, 
which may take values in, e.g., the set of real numbers or some real Lie group. We denote 
the integration manifold  as $M$ and let $\int_M\mathcal{D}x$ indicate a suitable 
integration over all possible field configurations of $x$. Expectation values of operators 
$\obs[x]$, which we shall henceforth simply refer to as observables, are then defined as
\begin{equation}
    \langle\obs\rangle := \int_M \mathcal{D}x\,\obs[x]\rho[x]\;.
\end{equation}

The idea behind the complex Langevin approach is to avoid having to deal with a complex 
weight $\rho[x]$ by analytically continuing the action and observables in $x$ and instead
consider field configurations $z$ of complexified degrees of freedom. For this to work, 
both $S$ and $\obs$ are required to be holomorphic. We then denote the 
complexified manifold, on which the components of $z$ are defined, as $M^c$. As we shall 
discuss below, the complex Langevin evolution gives rise to a real probability density 
$P[z]$, from which it allows one to draw samples, thus avoiding the sign problem. The 
crucial point is the question whether or not $P[z]$ is equivalent to $\rho[x]$ in the sense 
that
\begin{equation}\label{eq:equivalence}
    \int_{M^c} \mathcal{D}z\,\obs[z]P[z] \overset{?}{=}
    \int_M \mathcal{D}x\,\obs[x]\rho[x]\;,
\end{equation}
or, in other words, whether the expectation values computed in a complex Langevin 
simulation reproduce the correct values $\langle\obs\rangle$. As we shall see, this 
equivalence can -- in general -- not be guaranteed. 

\subsection{Basic formalism}
The complex Langevin approach makes use of the evolution of the complexified degrees of 
freedom $z$ in an auxiliary time direction, the so-called Langevin time $\tau$. In the 
following, we consider a single complex degree of freedom, $z=x+\ii y$. The $\tau$-
evolution of $z$ is then given by the following (complex) Langevin equation:
\begin{equation}\label{eq:cle}
    dz(\tau) = D(z)d\tau + \sqrt{K}dw(\tau)\;, \quad D(z) = -KS'(z)\;.
\end{equation}
It is a stochastic differential equation in which $dw$ denotes the increment of a standard 
Wiener process such that
\begin{equation}
    \langle dw(\tau)\rangle = 0\;, \quad \langle dw^2(\tau)\rangle = 2d\tau
\end{equation}
and $K\in\C$ is a so-called kernel whose purpose is to introduce a nontrivial 
diffusion into the Langevin equation. We have assumed it to be independent of $z$ (and 
$\tau$) here for simplicity. 

The stochastic evolution of $z$ in $\tau$ gives rise to a probability density
$P(x,y;\tau)$ in the complex plane, whose time-evolution is, in turn, given by the
Fokker--Planck equation
\begin{equation}\label{eq:fpe}
    \frac{\partial P(x,y;\tau)}{\partial \tau} = L^TP(x,y;\tau)\;,
\end{equation}
where we have defined the (real) Fokker--Planck operator 
$L^T:=\partial_x^2-\partial_x\re\,D-\partial_y\im\,D$\;.
It is a well-known fact that the complexification from $M$ to $M^c$ leads to a certain 
loss of mathematical rigor within the approach. In fact, while the convergence of the 
Fokker--Planck equation to the desired equilibrium distribution can be shown under rather 
mild assumptions in the case of a real action, the situation is less clear for 
complex actions. Additionally, in real Langevin simulations the introduction of a (real) kernel
does not alter the equilibrium distribution, whereas for complex actions a complex kernel
generally does.

One problem that may arise is the emergence of so-called runaway trajectories. The existence 
of these (classical) trajectories can cause the $\tau$-evolution of $z$ in the complex 
plane to wander far away from the real axis, which might bias sample averages and, if not 
accounted for, even lead to divergent simulations. As an example, consider the theory 
$S(z)=z^4$. Starting from an initial value $z_0=\ii y_0$ (with arbitrary $y_0\neq0$),
the classical evolution (i.e., the evolution without the noise term $dw$ in \eqref{eq:cle})
with a trivial kernel, $K=1$, will diverge. The noise term will, in general, kick the stochastic 
evolution off such a runaway trajectory, but, in practice, results might still be biased if a
simulation spends too much (Langevin) time far away from the real line, where discretization
effects might become significant. As a partial cure, one nowadays 
commonly uses an adaptively controlled Langevin time step \cite{AJS10} in the discretized 
evolution equations, which in many cases solves the runaway problem. Either way, runaway 
trajectories can at the very least be detected, namely by monitoring the distance of 
$z(\tau)$ to the real axis in a suitable way. For instance, in the above example one could 
simply keep track of $\im\, z$ in order to see whether runaway trajectories cause any 
problems.

The second major problem of complex Langevin simulations cannot be detected in such a 
straightforward way. Namely, the equilibrium distribution obtained from the Fokker--Planck 
evolution \eqref{eq:fpe}, 
\begin{equation}
    P(z):=\lim_{\tau\to\infty}P(x,y;\tau)\;,
\end{equation}
might not reproduce the correct expectation values $\langle\obs\rangle$, i.e., it might 
violate \eqref{eq:equivalence}. We shall devote the next subsection to a more detailed 
discussion of this problem.

\subsection{Boundary terms}
First of all, one might ask whether a given $\rho(z)$ actually admits the existence of any 
$P(z)$ satisfying \eqref{eq:equivalence} at all. As was shown in \cite{Sal97,Wei02,Sal07}, 
however, this is indeed the case for rather general $\rho(z)$. Whether complex Langevin 
simulations are capable of producing such a $P(z)$ is, of course, an entirely 
different question. A formal proof of correctness of the complex Langevin approach, i.e., 
of \eqref{eq:equivalence}, was given in \cite{AJS11}. This proof, however, relies on the 
sufficiently fast decay of certain quantities, e.g., the probability density 
$P(x,y;\tau)$, in the complex plane, such that one may perform an integration by parts 
without the appearance of boundary terms. However, it turns out that such boundary terms 
are, in fact, present quite generally.

A major advance was the realization that the presence of boundary terms in a complex 
Langevin simulation can be detected by measuring certain auxiliary observables 
\cite{SSS19,SSS20}. In particular, the boundary term expectation value for an observable 
$\obs$ can be defined for a single degree of freedom $z$ as
\begin{equation}\label{eq:boundary_terms}
    \B_\obs(Y) := \langle\Theta(Y-\mathcal{N}(z))L_c\obs(z)\rangle\;,
\end{equation}
where we have introduced $L_c:=(\partial_z-S'(z))K\partial_z$\;, $\mathcal{N}(z)$ denotes 
a suitable norm of $z$ and the Heavyside $\Theta$ function is used to control the spread 
of the distribution of $z$ in the complex plane. In practice, one looks for a plateau of 
$\B_\obs(Y)$ in the cutoff $Y$ and extrapolates to $Y\to\infty$, as $B_\obs$ tends to be 
very noisy when evaluated directly at $Y=\infty$. If such a plateau occurs at a 
nonvanishing value of $B_\obs$, one concludes that boundary terms are present and hence 
the assumptions underlying the proof in \cite{AJS11} do not hold. This, in turn, implies 
that the complex Langevin estimate for the expectation value of $\obs$, which we 
henceforth denote as $\langle\obs\rangle_\CL$, does not agree with the correct 
result $\langle\obs\rangle$ and should therefore be discarded. Indeed, the study of 
boundary terms to assess the correctness of complex Langevin results has become a standard 
by now \cite{LS23,ALR23,ARS24,HS24}. We mention that smaller boundary terms are usually 
associated with a smaller deviation between $\langle\obs\rangle_{\CL}$ and 
$\langle\obs\rangle$.

It is important to note that, while the presence of boundary terms can be safely 
interpreted as wrong convergence (if the stochastic evolution converges at all),
i.e., the probability density $P(z)$ emerging from a 
simulation not satisfying \eqref{eq:equivalence}, the converse is not necessarily true. 
Indeed, the absence of boundary terms is only a necessary requirement for correct 
convergence, not a sufficient one. A simple counterexample is discussed in \cite{MHS24p}. This
issue is particularly relevant in the presence of a kernel, and in the remainder of this 
work we shall discuss it in more detail.

%% file: cycles.tex
\subsection{General considerations}
In \cite{Sal93}, the wrong convergence of complex Langevin simulations was traced back to 
the fact that the (complex) Fokker--Planck equation in general features multiple solutions in the 
space of distributions, all of which may be sampled in a simulation. This observation was 
later formalized in \cite{SS19} in the form of a theorem that provides a sound explanation 
for incorrect convergence despite the absence of boundary terms in one-dimensional theories.

Concretely, \cite{SS19} considers integration paths in the complex plane that either 
connect two distinct zeros of $\rho(z)=e^{-S(z)}$ (with $S(z)$ denoting the action of a 
theory analytically continued to complex arguments) or are closed and noncontractible. In 
the following, we shall refer to such paths as \emph{integration cycles}\footnote{More 
rigorously, an integration cycle is an equivalence class of such paths or, in the language 
of algebraic topology, a so-called relative homology cycle \cite{Wit11}. In this work, 
however, we shall refrain from using that terminology.}, thereby borrowing 
terminology used in \cite{Wit11,Wit10}. The main statement of \cite{SS19} is that any 
linear functional that satisfies the Dyson--Schwinger equations of a one-dimensional theory 
with an action $S$ on a suitable space of test functions is given by a linear 
combination of integrals along such integration cycles; see also \cite{GGG96,GP09,GG10} for 
previous studies on the role of integration cycles in the context of Dyson--Schwinger 
equations. The relevance of this theorem for the complex Langevin approach can be appreciated 
by first realizing that the complex Langevin estimate $\langle\obs\rangle_\CL$
for the expectation value of an observable $\obs(z)$ is precisely such a linear functional, 
defined via the left-hand side of \eqref{eq:equivalence}. Second, the absence of 
boundary terms in a complex Langevin simulation is assumed to imply the validity of the 
aforementioned Dyson--Schwinger equations. 

Defining the expectation value of an observable $\obs$ along some integration cycle $\gamma_i$ as
\begin{equation}\label{eq:cycle_integral_1d}
    \langle\obs\rangle_{\gamma_i} := 
        \frac{\int_{\gamma_i}dz\, \obs(z) e^{-S(z)}}{\int_{\gamma_i}dz\, e^{-S(z)}}\;,
\end{equation}
the theorem thus states that $\langle\obs\rangle_\CL$, obtained in a simulation 
with vanishing boundary terms, is a linear combination of the 
$\langle\obs\rangle_{\gamma_i}$, i.e.,
\begin{equation}\label{eq:salcedo_seiler}
    \langle\obs\rangle_\CL = \sum_{i=1}^{N_\gamma}
        a_i\langle\obs\rangle_{\gamma_i}\;.
\end{equation}
Here, $a_i$ are complex coefficients, which, importantly, are the same for all observables 
$\obs$, and $N_\gamma$ denotes the number of linearly independent integration cycles of the 
theory. Notice that \eqref{eq:salcedo_seiler} immediately implies 
\begin{equation}\label{eq:constraint}
    \sum_{i=1}^{N_\gamma} a_i = 1\;,
\end{equation}
which can be used as a consistency check. 

In practice, one is usually interested in computing the average along some given 
integration cycle, which could be the real line for instance. Defining $\gamma_1$ to 
denote this integration cycle of interest, \eqref{eq:salcedo_seiler} states that in 
general a complex Langevin simulation may -- even in the absence of boundary terms -- not 
produce the desired result $\langle\obs\rangle_{\gamma_1}$, but also contributions from 
the other integration cycles $\gamma_{i\neq1}$. The obvious question that arises is in 
which situations one may find $a_i=\delta_{i1}$, such that 
$\langle\obs\rangle_\CL=\langle\obs\rangle_{\gamma_1}$. In this work, we aim at 
shedding some light on this question. In particular, we demonstrate that the coefficients 
$a_i$ can -- in principle -- be controlled by the kernel $K$ in \eqref{eq:cle}.
For the simple theories considered here, this can even be done in a rather systematic way. 
Indeed, since for these models $\langle\obs\rangle_{\gamma_i}$ can be computed directly 
for a sufficiently large set of observables $\obs$, we 
may simply extract the $a_i$ from \eqref{eq:salcedo_seiler} via a least-squares fit.
For more complicated theories, this is, of course, not possible. In fact, in general one 
does not even know what the independent integration cycles $\gamma_i$ are, nor could one 
hope to compute $\langle\obs\rangle_{\gamma_i}$ directly.

We emphasize that the theorem \eqref{eq:salcedo_seiler} was proven only for a single degree of 
freedom. While the concept of integration cycles can be extended to an arbitrary number of 
dimensions $d$ in a straightforward way, the validity of \eqref{eq:salcedo_seiler} beyond 
$d=1$ has not yet been established. In the remainder of this work, we thus 
investigate the theorem in both one and two dimensions from a numerical point of view. 
Before doing so, however, we first illustrate the main ideas behind integration cycles, 
including their linear independence, in a one-dimensional toy model. Moreover, we present 
an algorithm for counting the number of independent integration cycles in a certain class 
of theories with an arbitrary number of degrees of freedom in \cref{app:counting_cycles}.

\begin{figure}[t]
    \centering
    \includegraphics[width=0.8\linewidth]{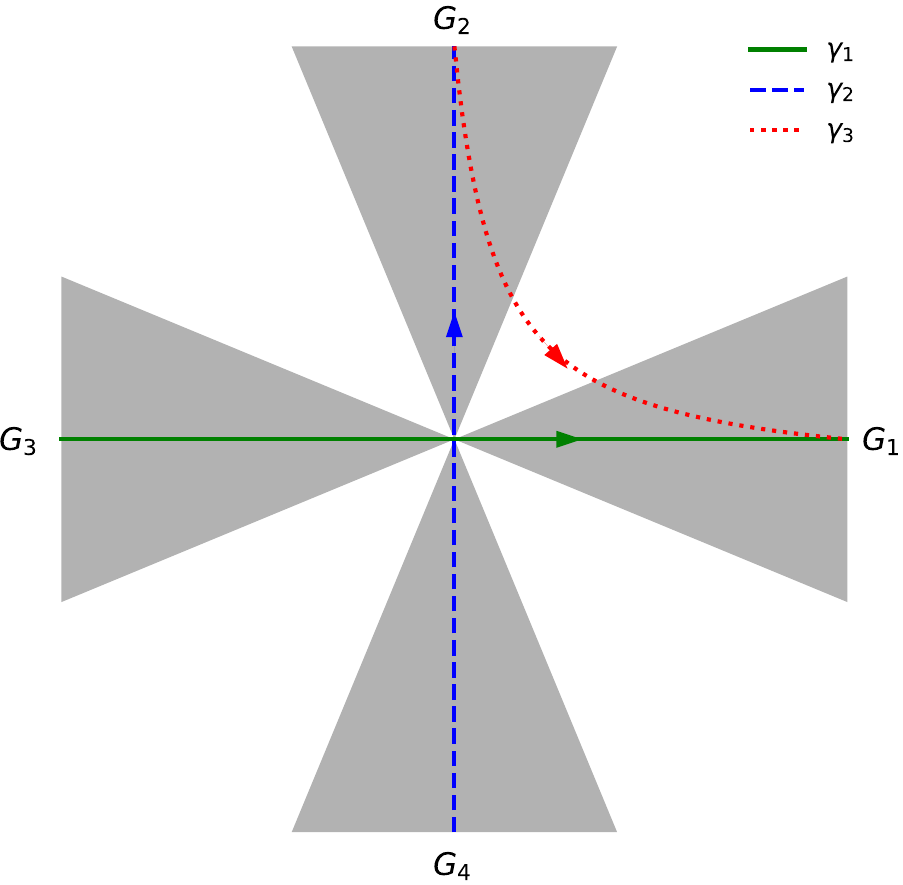}
    \caption{Topography of $e^{-S(z)}$ in the complex $z$ plane. The shaded areas are the 
             regions $G_i$ defined in \eqref{eq:regions_1d}. Possible realizations of the three 
             independent integration cycles in \eqref{eq:basis_cycles_1d} are shown as full, 
             dashed, and dotted lines, respectively.}
    \label{fig:cycles_1d}
\end{figure}

\subsection{One-dimensional example}\label{sec:cycles_1d}
Let us consider the theory defined by $S(z)=z^4$, with $z\in\C$, as an example. The density
$\rho(z)=e^{-S(z)}$ vanishes in four disjoint regions as $\vert z\vert\to\infty$. Namely, 
introducing polar coordinates $z=re^{\ii\theta}$, $\rho(z)$ approaches zero in the limit 
$r\to\infty$ if and only if $\theta\in G_i$, where
\begin{align}\label{eq:regions_1d}
    \begin{aligned}
        G_1 = \left(-\frac{\pi}{8},\frac{\pi}{8}\right)\;, \quad
        &G_2 = \left(\frac{3\pi}{8},\frac{5\pi}{8}\right)\;, \\
        G_3 = \left(\frac{7\pi}{8},\pi\right]\cup\left(-\pi, -\frac{7\pi}{8}\right)\;, 
        \quad
        &G_4 = \left(-\frac{5\pi}{8},-\frac{3\pi}{8}\right)\;,
    \end{aligned}
\end{align}
see \cref{fig:cycles_1d}. In particular, each of these regions contains one of the limits
$z\to\pm\infty$ and $z\to\pm\ii\infty$, respectively. 
Since $\rho(z)$ has no singularities, the set of integration cycles is made up of those 
paths that connect two different such regions. Moreover, since we consider only 
holomorphic observables, due to Cauchy's theorem an integration cycle is uniquely 
specified by the two regions $G_i$ it connects, i.e., it neither depends on the precise 
shape of the integration path nor on which exact point within the $G_i$ (at 
$\vert z\vert =\infty$) it starts and ends on. We thus introduce the notation $\gamma_{ij}$ 
for the unique integration cycle that connects regions $G_i$ and $G_j$ ($i\neq j$) at infinity.

Hence, in the model $S(z)=z^4$ there is a total of six (twelve if we count different 
orientations) integration cycles, only $N_\gamma=3$ of which, however, are linearly 
independent. This is due to the following identity:
\begin{equation}\label{eq:line_integrals}
    \int_{\gamma_{ij}}dz + \int_{\gamma_{jk}} dz = \int_{\gamma_{ik}}dz \;.
\end{equation}
Thus, a possible basis of integration cycles for this theory is given by
\begin{equation}\label{eq:basis_cycles_1d}
    \gamma_1=\gamma_{31}\;, \quad \gamma_2=\gamma_{42}\;, \quad \gamma_3=\gamma_{21}\;,
\end{equation}
which are homotopic (homologous) to the real line, the imaginary line and an integration path 
connecting $\ii\infty$ and $\infty$, respectively. These integration contours are 
depicted in \cref{fig:cycles_1d} as well. 

Now consider the theory $S(z)=\lambda z^4$ with $\lambda\in\C$. In this case, the regions 
$G_i$ of vanishing $\rho(z)$ result from those given in \eqref{eq:regions_1d} after a rotation 
by $\arg\sqrt[4]{1/\lambda}$. With these, one may again define a basis of integration cycles 
via \eqref{eq:basis_cycles_1d}. Unsurprisingly, this implies that the real line defines a 
valid integration cycle only as long as $\re\,\lambda>0$, which has important consequences 
for the convergence of complex Langevin simulations. These statements shall be discussed in 
more detail in \cref{sec:results}

The generalization to higher-order polynomial actions
\begin{equation}
    S(z) = \sum_{n=1}^{N_0}\lambda_nz^n
\end{equation}
with $\lambda_{N_0}\neq0$ is now straightforward. Namely, there are $N_0$ regions 
$G_i$ and thus, due to \eqref{eq:line_integrals}, $N_0-1$ linearly independent integration 
cycles, which may be found via a similar procedure as outlined above. The generalization to 
more than one degree of freedom is more involved and discussed in 
\cref{app:counting_cycles}.

%% file: simulation_setup.tex
In this section, we discuss details of the numerical simulations performed in order to 
obtain the results presented in \cref{sec:results}. In accordance with the FAIR 
guiding principles \cite{FAIR16}, we provide access to our simulation data online \cite{data}
and furthermore publish our analysis scripts \cite{scripts}. Our simulation code
is also available upon request.

\subsection{Discrete Langevin evolution}
For our one- and two-dimensional simulations, we employ a discretized version of the 
Langevin equation \eqref{eq:cle}. We use the following improved update 
\cite{Ukawa:1985hr}
of a configuration\footnote{Here and in the following 
there is no implicit summation over repeated indices.} $z_i$:
\begin{equation}\label{eq:discrete_cle}
    z_i\to z_i - \eps \frac{D_i(z)+D_i(z')}{2} + \sqrt{\eps K_i}\eta_i\;,
\end{equation}
where $i$ labels the degrees of freedom ($i=1,\dots,d$), the drift $D_i(z)$ is defined as 
\begin{equation}
    D_i(z) := K_i \frac{\partial S(z)}{\partial z_i}
\end{equation}
and we have introduced the auxiliary intermediate configuration
\begin{equation}\label{eq:intermediate_configuration}
    z_i' := z_i -\eps D_i(z) + \sqrt{\eps K_i}\eta_i\;.
\end{equation}
In the above equations, $\eps$ denotes the discretized Langevin-time step size and 
the $\eta_i$ are independent random variables drawn from a Gaussian distribution (with mean 
$0$ and variance $2$) for every update step. To generate the Gaussian random numbers, we 
employ a Combined Multiple Recursive Generator (CMRG) as described in \cite{LSC02}, 
combined with a standard Box--Muller transformation \cite{BM58}. Note that for $d>1$, the 
kernel, chosen to be independent of $z_i$ here, can be a full matrix 
$K_{ij}$ in general. However, we have chosen it to be diagonal, $K_{ij}=K_i\delta_{ij}$, in 
this work for simplicity. The more general case of $z_i$-dependent and/or full matrix 
kernels will be discussed elsewhere.

\subsection{Adaptive-step-size algorithm}
In order to mitigate the influence of runaway trajectories on our simulations, we determine 
the step size $\eps$ adaptively in the following way \cite{AJS10}:
\begin{enumerate}
    \item Consider the step size $\eps$ from the previous update.
    \item\label{step:adaptive_stepsize} Propose a $z_i'$ via 
    \eqref{eq:intermediate_configuration}, compute
    \begin{equation}
            D_{\max} := \max_i \left\vert\frac{D_i(z)+D_i(z')}{2}\right\vert
    \end{equation}
    and check whether
    \begin{equation}\label{eq:adaptive_stepsize}
        \frac{\mathcal{D}}{2} \leq \eps D_{\max} \leq 2\mathcal{D}\;,
    \end{equation}
    where $\mathcal{D}$ is some suitably chosen reference value. 
    \item If $\eps D_{\max}$ lies outside that interval, multiply $\eps$ 
    with appropriate powers of $2$ until \eqref{eq:adaptive_stepsize} holds. If the 
    step size grows larger than a given upper limit $\eps_{\max}$ in the process, set 
    $\eps=\eps_{\max}$. 
    \item Go back to step \ref{step:adaptive_stepsize} with the new value of $\eps$ and 
    repeat these steps until \eqref{eq:adaptive_stepsize} is satisfied at first try.
\end{enumerate}
Once we have determined the step size in this way, we insert it into \eqref{eq:discrete_cle} 
to proceed to the next configuration and move forward in Langevin 
time by $\eps$. We perform this procedure for every update step until the 
simulation reaches a certain maximum Langevin time $\tau_{\max}$, at which it terminates.

\subsection{Parallelization}
In order to reduce statistical uncertainties as much as possible, we utilize the massive 
parallelization capabilities of modern GPUs via the \verb|CUDA| application programming 
interface \cite{cuda}. In fact, due to the sheer simplicity of the models considered here, 
we may actually let one entire simulation, in the way outlined above, run on each individual 
\verb|CUDA| thread we launch. Naturally, each of these simulations starts from a different 
random initial configuration and has an independent random-number seed. For a single 
simulation run, we launch $\Nsim=\mathcal{O}(10^4)$ \verb|CUDA| threads and we 
typically perform $\Nruns=\mathcal{O}(10^2)$ such runs for each parameter value we are 
interested in. Notice that we distinguish between \emph{simulations}, each of which is 
performed by a single \verb|CUDA| thread, and \emph{runs}, which consist of $\Nsim$ 
such simulations each.

Within each run, every simulation terminates at the same $\taumax$ and we also ensure 
that $\taumax$ is the same across runs. Moreover, measurements, which are discussed in 
detail in the next subsection, are performed whenever each simulation in a run has 
progressed by a Langevin time of $\taumeas$, where $\taumeas$ is chosen to agree between 
different runs as well. We also monitor whether individual simulations diverge, but, 
due to the employed adaptive step size algorithm, this barely ever happens and its 
effect can be neglected entirely. However, what may occur on rare occasions is 
that one or more simulations are slowed down significantly by the adaptive step size 
algorithm, which, in turn, affects the entire run due to the synchronization of threads 
before and after measurements.

\subsection{Measurement}\label{sec:measurement}
We do not store the full set of configurations produced by each simulation of a given run, 
as this would exhaust the available storage rather quickly. Instead, we measure 
observables as follows: Whenever a simulation has run for a (Langevin) time $\taumeas$ since the 
last measurement, it computes all observables $\obs_i$ of interest, which are -- after 
synchronization -- averaged over all simulations within the run, i.e., over all \verb|CUDA| 
threads involved. Only these (thread-)averages are then written to file in every 
measurement step. The main disadvantage of this approach is that one loses information 
about the configurations produced by individual simulations, as well as their 
correlations. In order to nonetheless gain some insight into the latter, as well as into 
the thermalization properties of the simulations, we do monitor the timelines of a handful 
of individual simulations every once in a while. Note that our strategy does not 
allow us to perform measurements of new observables 
without re-generating the ensembles from scratch.

\subsubsection{Regular observables}\label{sec:measurement_regular_observables}
An estimator for the expectation value of an observable, $\langle\obs_i\rangle_{\CL}$, is 
computed in the following way: For each run, the (thread-averaged) timeline of 
$\obs_i$ is read and all measurements up to a Langevin time of $\tautherm$ are discarded 
to account for thermalization. The thermalized timelines are then averaged over Langevin 
time, leaving us with a total of $\Nruns$ (uncorrelated) data points for each observable. 
Finally, the estimator $\langle\obs_i\rangle_{\CL}$ is computed as the average of these 
data points over all runs, while its statistical uncertainty is their standard deviation 
divided by $\sqrt{\Nruns}$.

\subsubsection{Boundary terms and histograms}\label{sec:measurement_boundary_terms_histograms}
The computation of boundary terms is somewhat trickier, as it has do be done online due 
to the $\Theta$ function in \eqref{eq:boundary_terms}, information of which is lost after 
the thread-averaging. We first fix (for all simulations 
in all runs) a range $[Y_{\min},Y_{\max})$ for the cutoff $Y$, which we discretize into 
$N_Y$ bins logarithmically. After the initial wait time $\tautherm$ to ensure 
thermalization, boundary terms for each observable $\obs_i$ are computed during every 
measurement step as follows: In each simulation, we compute the norm $\mathcal{N}(z)$ in 
\eqref{eq:boundary_terms}, defined here as 
$\mathcal{N}(z):=\max_i\left\vert z_i\right\vert$, to determine the corresponding $Y$ 
bin and add the quantity $L_c\obs_i(z)$ to that bin. After the run has finished, the bins 
are normalized by the number of thermalized measurements, 
$\Nmeas=(\taumax-\tautherm)/\taumeas$, 
(which is an integer with our choice of parameters) as well as the 
number of simulations  $\Nsim$ in that run. The $\Theta$ function in 
\eqref{eq:boundary_terms} is then implemented by adding each $Y$ bin to all bins that 
correspond to larger values of $Y$.
This leaves us with one value of $B_{\obs_i}$ per bin for each run, from which we 
obtain estimators for the expectation value and uncertainty in the same way as for the 
regular observables. We also note that histograms of the configurations may be computed in a 
similar fashion.

\subsubsection{Integration cycle coefficients}\label{sec:measurement_coefficients}
In order to determine the coefficients $a_i$ in \eqref{eq:salcedo_seiler}, we employ the 
following strategy: First of all, consider the output of a single simulation run, which 
consists of $\Nmeas$ thermalized measurements for each (thread-averaged) observable. The
$\tau$-average of these quantities thus already provides us with a reasonable estimate 
$\overline{\obs}_i$ for $\langle\obs_i\rangle_\CL$ that we may use in the left-hand-side 
of \eqref{eq:salcedo_seiler} to extract the $a_i$ via a least-squares fit. Defining the 
number of measured observables as $N_\obs$, we therefore require $N_\obs\geq N_\gamma$. 
Importantly, there are nonnegligible correlations between the observables that we take 
into account by computing their covariance matrix. In order to do so in a well-defined 
way, we first split all observables into their real and imaginary parts and consider those 
as independent (real) observables. Then, \eqref{eq:salcedo_seiler} can (with the replacement 
$\langle\obs_i\rangle_\CL\to\overline{\obs}_i$) be written as the matrix equation
\begin{equation}
    \boldsymbol{y} = X\boldsymbol{\beta}\;,
\end{equation}
where $\boldsymbol{y}$ and $\boldsymbol{\beta}$ are real vectors with $2N_\obs$ and 
$2N_\gamma$ components, respectively, which are defined as
\begin{align}
    \begin{aligned}
        y_{2i-1} = \re\,\overline{\obs}_i\;, 
            \quad y_{2i} = \im\,\overline{\obs}_i\;, 
            \quad 1\leq i\leq N_\obs\;,\\
        \beta_{2j-1} = \re\,a_j\;, 
            \quad \beta_{2j} = \im\,a_j\;, 
            \quad 1\leq j\leq N_\gamma\;,
    \end{aligned}
\end{align}
and $X$ is the $(2N_\obs\times2N_\gamma)$-matrix resulting from 
$\langle\obs_i\rangle_{\gamma_j}$ via
\begin{align}
    \begin{aligned}
        X_{2i-1,2j-1} &= \re\langle\obs_i\rangle_{\gamma_j}\;, \quad 
        X_{2i-1,2j} = -\im\langle\obs_i\rangle_{\gamma_j}\;, \\
        X_{2i,2j-1\phantom{-1}} &= \im\langle\obs_i\rangle_{\gamma_j}\;, \quad
        X_{2i,2j\phantom{-1}} = \re\langle\obs_i\rangle_{\gamma_j}\;.
    \end{aligned}
\end{align}
Then, denoting the covariance matrix of $\boldsymbol{y}$ (computed as an average over 
$\tau$ and normalized by $1/\sqrt{\Nmeas}$) by $\Sigma$, the least-squares estimator $\boldsymbol{\hat{\beta}}$ is obtained by 
minimizing
\begin{equation}\label{eq:chi_squared}
    \chi^2 := 
    (\boldsymbol{y}-X\boldsymbol{\beta})^T\Sigma^{-1}(\boldsymbol{y}-X\boldsymbol{\beta})
\end{equation}
and reads
\begin{equation}\label{eq:least_squares}
    \boldsymbol{\hat{\beta}} = (X^T\Sigma^{-1}X)^{-1}X^T\Sigma^{-1}\boldsymbol{y}\;.
\end{equation}
Thus, after deciding on a suitable set of observables, we may compute 
$\boldsymbol{\hat{\beta}}$ for every run and obtain a final estimate for the coefficients
$a_i$ and their uncertainties by averaging over runs and computing the standard deviation
(divided by $\sqrt{\Nruns}$), as before.

\subsection{Simulation parameters}
For the runs discussed in this work, we fix the maximum Langevin time per simulation to 
$\taumax=1000$ and we store (thread-averaged) measurements after every $\taumeas=0.1$. The 
step size is initialized with a value of $\eps=10^{-5}$ and not allowed to grow beyond 
that value, i.e.,  $\eps\leq\eps_{\max}=10^{-5}$. The reference value in 
\eqref{eq:adaptive_stepsize} is chosen to be $\mathcal{D}=10^{-2}$. For each run, we 
simultaneously launch $\Nsim=2^{13}$ \verb|CUDA| threads on $2^8$ thread blocks and we 
perform between $\Nruns=100$ and $\Nruns=200$ runs for each parameter value of interest. We 
compute averages after discarding measurements up to a thermalization time of 
$\tautherm=50$, after which we also start measuring boundary terms. The range of the 
cutoff $Y$ is set to $[Y_{\min},Y_{\max})=[10^{-5},40)$ and partitioned into $N_Y=200$ bins 
logarithmically. For a few runs, we moreover compute histograms, for which the same number 
of configurations is discarded. This is important because unthermalized configurations
might wrongly suggest that histograms have power-law tails when they actually
decay exponentially.

%% file: results.tex
Let us now turn to the discussion of our simulation results. The first main goal of this 
investigation is to check the validity of \eqref{eq:salcedo_seiler} numerically, both in one 
dimension, where it was proven in \cite{SS19}, and in two dimensions, where it is only a 
conjecture at this point. Second, we also aim at establishing a relation between the 
kernel $K$ in \eqref{eq:cle} and the coefficients $a_i$ in \eqref{eq:salcedo_seiler}.

\subsection{One dimension}\label{sec:results_1d}

To begin with, let us consider the model
\begin{equation}\label{eq:quartic_1d}
    S(z) = \frac{\lambda}{4}z^4\;,
\end{equation}
with a complex coefficient $\lambda$ that is parametrized by an integer $l$ as
\begin{equation}\label{eq:lambda}
    \lambda = e^{\ii\pi l/6}\;, \quad l\in \{-5,\dots,6\}.
\end{equation}
The effect of a kernel on complex Langevin simulations of this model was
first studied in \cite{OOS89} and previous results regarding the relation between a
kernel and the theorem \eqref{eq:salcedo_seiler} in the model can be found in \cite{MHS24p}. 

Our simulation setup was described in detail in \cref{sec:simulation_setup}. In particular, 
we use \eqref{eq:intermediate_configuration} and \eqref{eq:discrete_cle} for the discrete 
Langevin update step, where, as in \cite{OOS89,MHS24p}, we parametrize the kernel as
\begin{equation}\label{eq:kernel_1d}
    K = e^{\ii\pi m/24}\;, \quad m\in \{0,\dots,47\}\;.
\end{equation}
We are interested in computing monomial observables of the form $\langle z^n\rangle$, $n>0$,
which in the model \eqref{eq:quartic_1d} can be computed exactly. Indeed, one finds
\begin{subequations}\label{eq:exact_1d}
    \begin{align}\label{eq:exact_1d_even}
        \frac{\int_{-\infty}^\infty dz\,z^{2k}e^{-S(z)}}
             {\int_{-\infty}^\infty dz\,e^{-S(z)}} &= 
        \left(\frac{4}{\lambda}\right)^{\frac{k}{2}}\frac{\Gamma((2k+1)/4)}{\Gamma(1/4)}\;,\\
        \frac{\int_{-\infty}^\infty dz\,z^{2k+1}e^{-S(z)}}
             {\int_{-\infty}^\infty dz\,e^{-S(z)}} 
        &= 0\;,
    \end{align}
\end{subequations}
for even and odd powers of $z$, respectively, where $k$ is a nonnegative integer and 
$\Gamma$ denotes the usual gamma function. Notice that in \eqref{eq:exact_1d} we integrate 
over the real axis, such that we require $\re\,\lambda>0$ (or $\vert l\vert<3$) for the 
integrals to exist, but one may analytically continue the right-hand side to arbitrary values of $\lambda\neq0$. 
However, an analytic expression can have ambiguities due to cuts and is generally 
unavailable. Therefore, as described below, we define the analytic continuation of such
integrals by making use of integration cycles, avoiding any such ambiguities.

\begin{figure}[t]
    \centering
    \includegraphics[width=1\linewidth]{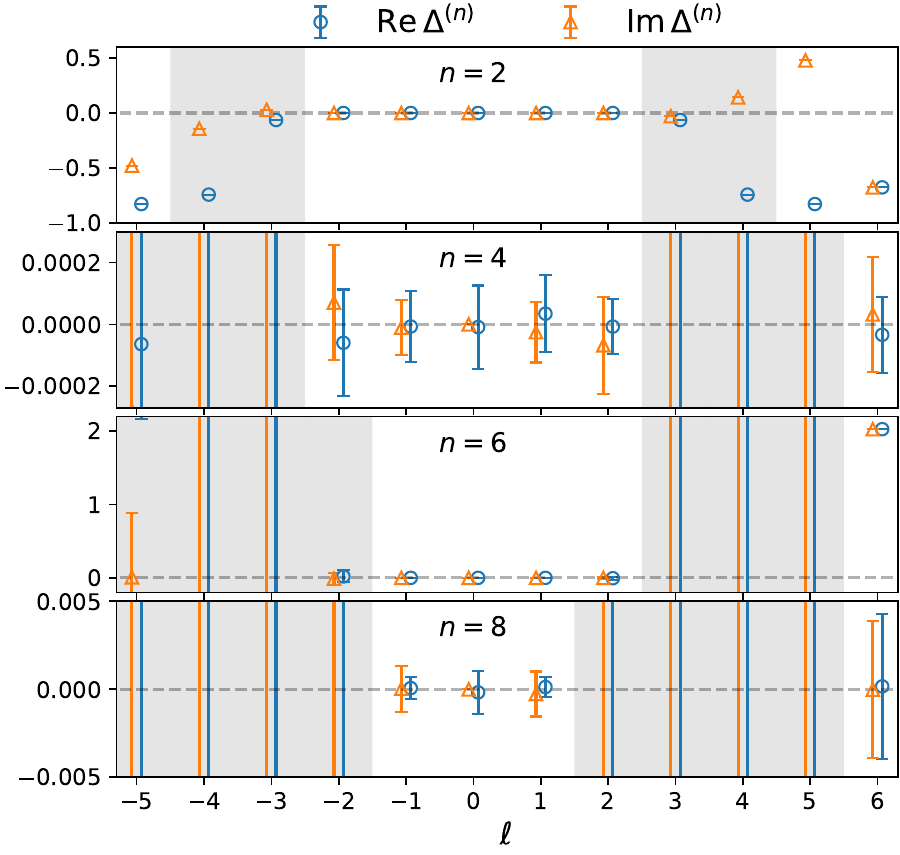}
    \caption{Real and imaginary parts of the difference           
             $\Delta^{(n)}$ in \eqref{eq:delta_1d} for even $n$ as a function of the integer 
             $l$ in \eqref{eq:lambda}. For values of $l$ within the shaded regions, the 
             corresponding observable has a nonzero boundary term. Notice the different 
             scales on the vertical axes; the axis limits have been chosen such that all 
             results without boundary terms fit within. The dashed horizontal lines indicate 
             $\Delta^{(n)}=0$ and the data points have been displaced horizontally for better 
             visibility. }
    \label{fig:observables_vs_l_1d_even}
\end{figure}

To this end, we first need to specify precisely what is meant by the generalization of the 
basis integration cycles \eqref{eq:basis_cycles_1d} for $\lambda\notin\R$ alluded to in 
\cref{sec:cycles_1d}. Concretely, we perform the variable transformation 
\begin{equation}\label{eq:variable_transformation}
    z\to\sqrt[4]{\lambda}z=:\xi\;.
\end{equation}
In the new variable $\xi$, the action \eqref{eq:quartic_1d} is real on the real axis for any 
choice of $\lambda$, such that the discussion in \cref{sec:cycles_1d} applies directly. In 
particular, we define the linearly independent integration cycles as in 
\eqref{eq:basis_cycles_1d}, except that the integration variable is now $\xi$ instead of 
$z$. This implies, for instance, that the integration cycle $\gamma_1$, which is homotopic 
to the real $\xi$ axis (such that we shall sometimes refer to it as the `real' integration 
cycle), is, in fact, not (homotopic to) a real contour in the original variable $z$ if 
$\lambda\notin\R$. With this definition and the notation introduced in 
\eqref{eq:cycle_integral_1d}, the right-hand side of \eqref{eq:exact_1d} is then simply 
given by $\langle z^n\rangle_{\gamma_1}$ without requiring any analytic continuation. This 
leads to a natural definition for the exact result of $\langle z^n\rangle$ as
\begin{equation}
    \langle z^n\rangle_{\mathrm{exact}} := 
        \langle z^n\rangle_{\gamma_1}\;.
\end{equation}
In the following, we shall examine under which conditions the result $\langle z^n\rangle_\CL$
obtained in a complex Langevin simulation coincides with 
$\langle z^n\rangle_{\mathrm{exact}}$.

\begin{figure}[t]
    \centering
    \includegraphics[width=\linewidth]{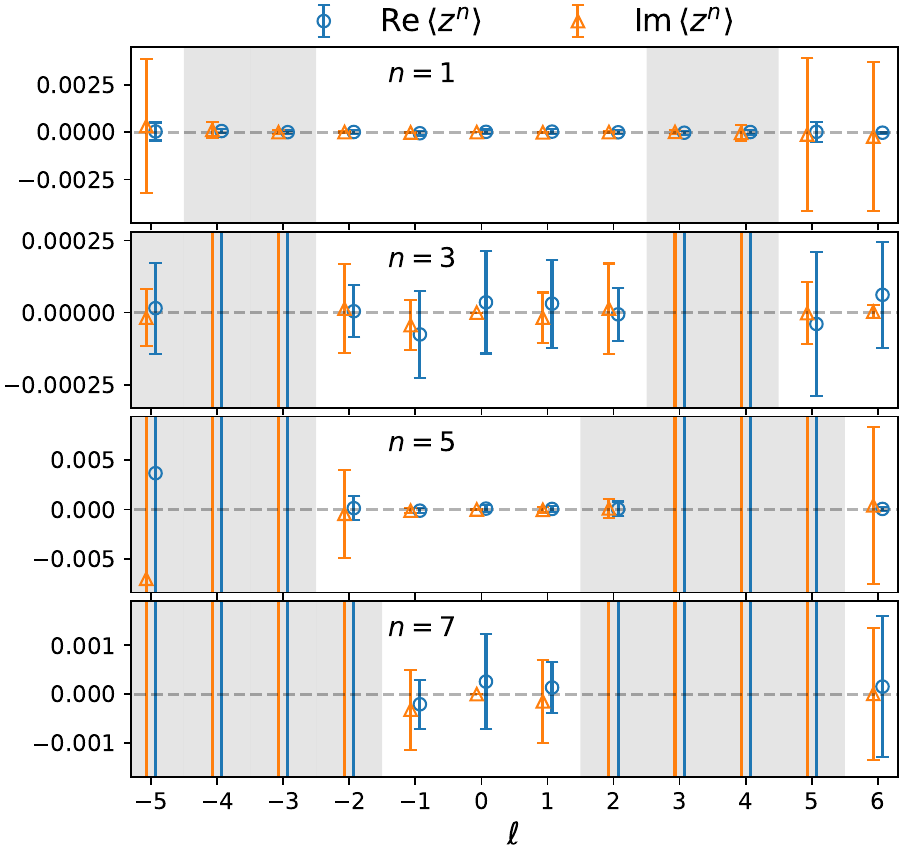}
    \caption{Real and imaginary parts of the observables $\langle z^n\rangle_{\mathrm{CL}}$ 
             for odd 
             $n$ as a function of the integer $l$ in \eqref{eq:lambda}. Otherwise analogous to 
             \cref{fig:observables_vs_l_1d_even} (note that $\Delta^{(n)}=\langle z^n\rangle_\CL$
             for odd $n$).}
    \label{fig:observables_vs_l_1d_odd}
\end{figure}

\subsubsection{Trivial kernel}
A comparison between $\langle z^2\rangle_{\mathrm{CL}}$ and 
$\langle z^2\rangle_{\mathrm{exact}}$, including a discussion of the corresponding boundary 
term $\B_{z^2}$, for $K=1$ (i.e., $m=0$) and different values of $\lambda$ was presented in 
\cite{MHS24p}. Here, we shall discuss this issue in more detail and consider higher 
powers in the observables as well. Concretely, we have measured $\langle z^n\rangle_{\mathrm{CL}}$ 
and $\B_{z^n}(Y)$ for $n\in\{1,\dots,8\}$. Defining the difference
\begin{equation}\label{eq:delta_1d}
    \Delta^{(n)}:=\langle z^n\rangle_{\mathrm{CL}}-\langle z^n\rangle_{\mathrm{exact}}\;,
\end{equation}
we compare simulation and exact results for various $\lambda$ and even values of $n$ in 
\cref{fig:observables_vs_l_1d_even}. For each $\lambda$ and $n$, we moreover look for plateaus 
in $Y$ of the corresponding boundary term. If such a plateau is assumed at a value of 
$\B_{z^n}$ that differs from zero by more than three times the statistical uncertainty or if there 
is no clear signal for such a plateau at all, we conclude that there is a nonvanishing boundary 
term and the result $\langle z^n\rangle_{\mathrm{CL}}$ must thus be incorrect. 
\footnote{Notice that this is not an exact procedure, with the result that the boundaries 
between regions with $\B_{z^n}=0$ and $\B_{z^n}\neq0$ are not sharp but might be subject to 
small variations. However, since smaller boundary terms are typically correlated with smaller 
deviations from $\langle z^n\rangle_{\mathrm{exact}}$, it makes little difference in practice 
whether boundary terms vanish exactly or are just very small.} For each observable, the values 
of $l$ for which we observe signals for boundary terms are indicated by the shaded regions in 
\cref{fig:observables_vs_l_1d_even}. Corresponding results for odd $n$, for which 
$\langle z^n\rangle_{\mathrm{exact}}=0$, are shown in \cref{fig:observables_vs_l_1d_odd}. We 
observe that generically the simulation results show good agreement with 
$\langle z^n\rangle_{\mathrm{exact}}$ for values of $l$ close to zero (or $\lambda$ close to 
the positive real axis), where there are no boundary terms. 

Obviously, the question whether $\B_{z^n}(Y)$ shows a plateau at a nonvanishing value may 
have a different answer for different $n$. The statement that boundary terms are absent, 
however, requires that such plateaus be consistent with zero for \emph{all} observables. In 
particular, as becomes clear from 
\cref{fig:observables_vs_l_1d_even,fig:observables_vs_l_1d_odd}, boundary terms might 
appear to be consistent with zero for some lower powers of $z$ but deviate from zero 
for higher powers. While one might argue that in such a case even the boundary 
terms for low orders should, in fact, be finite (but small and potentially hard to distinguish 
from zero), one concludes that it is advisable in general to consider a large set of 
observables covering a broad range of powers. This is particularly important since the theorem 
\eqref{eq:salcedo_seiler} relies on the absence of boundary terms for all observables. 

In the earlier study \cite{MHS24p}, only the boundary terms for $z^2$ were taken into account. 
Looking at \cref{fig:observables_vs_l_1d_even,fig:observables_vs_l_1d_odd}, however, this 
restriction can actually turn out to be misleading in some cases. For instance, the results of 
\cite{MHS24p} suggest that boundary terms are absent for $l=5$, while in reality they become 
significant only for $z^n$ with $n\geq4$. In such a case, one should thus not expect 
\eqref{eq:salcedo_seiler} to hold. In passing, we mention that, as seen in 
\cref{fig:observables_vs_l_1d_even,fig:observables_vs_l_1d_odd}, the statistical uncertainties 
tend to be much larger in the presence of boundary terms, especially for large $n$, reflecting 
the slow decay of the corresponding probability distributions in the complex plane. 

We stress that one should not be tempted at this point to interpret the presence of boundary 
terms as an indication for convergence to `incorrect' results, as one would in simulations of 
realistic theories. After all, we have not defined what the `correct' results should be -- all 
we do is compare with \eqref{eq:exact_1d}. In that respect, a nonzero boundary term $\B_{z^n}$ 
simply means that the associated probability distribution has slowly decaying tails. It is 
interesting to observe that the resulting large error bars actually in many cases allow for 
the interpretation that $\langle z^n\rangle_{\mathrm{CL}}$ could, in fact, be consistent with 
$\langle z^n\rangle_{\mathrm{exact}}$. However, these results are still unusable as they are 
entirely dominated by noise despite the large amount of statistics in our setup. In other 
cases, especially for low powers $n$, error bars are small but the results nonetheless 
disagree with $\langle z^n\rangle_{\mathrm{exact}}$. These are the two ways in which boundary 
terms can distinguish between good and bad results. 

The crucial point that we would like to emphasize, however, is that -- on its own -- the 
absence of boundary terms does not automatically imply that 
$\langle z^n\rangle_{\mathrm{CL}}=\langle z^n\rangle_{\mathrm{exact}}$. The data for $l=6$, 
for instance, serve as a counterexample, as there is a clear disagreement between the complex 
Langevin results and \eqref{eq:exact_1d_even} for even powers $n$, despite the boundary terms 
being consistent with zero for all $n$ considered. This observation shall now be explained as 
contributions from unwanted integration cycles $\gamma_{i\neq1}$.

In order to compute the coefficients $a_i$ in \eqref{eq:salcedo_seiler}, we follow the 
procedure outlined in \cref{sec:measurement}, where in the least-squares fit we use the 
$N_\gamma=3$ independent integration cycles \eqref{eq:basis_cycles_1d} and the following set 
of observables:
\begin{equation}\label{eq:observable_set_1d}
    \left\{\langle z\rangle, \langle z^2\rangle, \langle z^4\rangle, 
           \langle z^5\rangle, \langle z^6\rangle, \langle z^8\rangle
    \right\}\;.
\end{equation}
In fact, out of the eight observables we have measured, \eqref{eq:observable_set_1d} is the 
largest set we can use in the fit, since $\langle z^{4k+3}\rangle_{\gamma_i}=0$ for all $i$ 
and nonnegative integers $k$, such that the exact results for $n=3$ and $n=7$ are trivial and 
do not provide any usable information. We remark that for the fit we have, in fact, computed 
$\langle z^n\rangle_{\gamma_i}$ via numerical integration to be consistent with the
two-dimensional case discussed below.

For diagnostic reasons, we have compared fits obtained 
with different sets of observables, also including exponential observables of the form 
$\langle e^{\alpha z^k}\rangle$ (with $\alpha\in\C$ and $k$ a positive integer). We found that 
(in the cases where the fit provides any reasonable result at all), the results are more or 
less independent of the chosen set as long as the latter is sufficiently large. In particular, 
our investigations suggest that any set of five or more different observables $\obs$ with at 
least one nontrivial 
$\langle \obs\rangle_{\gamma_i}$ will provide consistent results. This is also the reason why we 
do not consider exponential observables in the remainder of this work. In order to assess the 
goodness of our fit results, we have analyzed the distribution of whitened residuals in a 
procedure that is outlined in \cref{app:chi_squared}. Whenever we say that a fit is good, we 
mean that it satisfies the criteria defined there. 

\begin{table*}[t]
    \centering
    \caption{Coefficients $a_i$ in the model \eqref{eq:quartic_1d} for $K=1$ and different values of $l$ 
             in \eqref{eq:lambda} for which boundary terms vanish, obtained from 
             \eqref{eq:salcedo_seiler} via the least squares fit procedure outlined in 
             \cref{sec:measurement_coefficients}, using the observables 
             \eqref{eq:observable_set_1d}. The results are rounded to the first significant 
             digit of the respective statistical uncertainties.}
    \renewcommand{\arraystretch}{1.3}
	\renewcommand{\tabcolsep}{5pt}
    \begin{tabular}{|c|c|c|c|}
    \hline
    $l$ & $a_1$ & $a_2$ & $a_3$\\
    \hline
    $-1$ & $1.00003(2)+0.000006(4)\ii$ & $0.000006(3)-0.00004(2)\ii$ & $-0.00005(2)+0.00003(1)\ii$\\
    $1$ & $1.00000(2)+0.000003(5)\ii$ & $0.000005(3)+0.00002(2)\ii$ & $\hphantom{-}0.00002(1)-0.00002(2)\ii$\\
    $6$ & $\hphantom{1}0.4999(4)-0.4999(4)\ii\hphantom{00}$ & $\hphantom{0}0.5001(4)+0.5001(4)\ii$ & $-0.00003(1)-0.0002(8)\ii\hphantom{0}$\\
    \hline
    \end{tabular}
    \label{tab:coefficients_1d}
\end{table*}

In general, we observe that the least-squares fit works well whenever there are no boundary 
terms but becomes unreliable when boundary terms are nonzero. Indeed, the presence of boundary 
terms is correlated with the covariance matrix $\Sigma$ in \eqref{eq:least_squares} becoming 
ill-conditioned, which results in its inversion and, by extension, the fit results becoming 
unstable. This is exactly the expected behavior, as \eqref{eq:salcedo_seiler} is valid only in 
the absence of boundary terms. Note that for $l=0$ all simulation trajectories are confined to 
the real axis after thermalization. In this case, the only integration cycle that is being 
sampled is $\gamma_1$, trivially, without the need of a fit, which is also consistent with
\eqref{eq:salcedo_seiler}. In fact, our approach does not allow us compute the coefficients 
$a_i$ in this case, as the imaginary parts of all observables vanish, rendering $\Sigma$ singular. 

The only values of $l\neq0$ for which boundary terms vanish with a trivial kernel are $l=\pm1,6$
and we present fit 
results for the coefficients $a_i$ for those values in \cref{tab:coefficients_1d}. We observe 
that $a_i\approx\delta_{i1}$ as long as $\lambda$ is sufficiently close to the positive real 
axis. In other words, for $\vert l\vert<2$ only the real integration cycle contributes 
significantly to our simulations with $K=1$, such that the latter reproduce 
$\langle z^n\rangle_{\mathrm{exact}}$, in accordance with 
\cref{fig:observables_vs_l_1d_even,fig:observables_vs_l_1d_odd}. For $l=6$, on the other hand,
$\gamma_1$ and $\gamma_2$ both contribute equally, explaining the deviations in 
\cref{fig:observables_vs_l_1d_even,fig:observables_vs_l_1d_odd} despite the absence of 
boundary terms. Importantly, the consistency condition \eqref{eq:constraint} is fulfilled for 
all values of $l$ in \cref{tab:coefficients_1d}. We note that the fact that 
$\langle z^{4k}\rangle_\CL\approx\langle z^{4k}\rangle_\mathrm{exact}$ for positive integers 
$k$ at $l=6$ is because $\langle z^{4k}\rangle_{\gamma_1}=\langle z^{4k}\rangle_{\gamma_2}$ 
due to symmetry. Also notice that $a_3$ is vanishingly small for all $l$ in 
\cref{tab:coefficients_1d}, indicating that the third integration cycle $\gamma_3$ is not 
sampled at all in our runs, which also explains the agreement between simulation and exact 
results for odd $n$, as 
$\langle z^{2k+1}\rangle_{\gamma_{1}}=\langle z^{2k+1}\rangle_{\gamma_{2}}=0$ for nonnegative 
integers $k$. This fact shall be discussed in more detail below. 

We conclude that all of our results presented so far are in excellent agreement with the 
theorem \eqref{eq:salcedo_seiler}. In particular, we observe that as long as boundary terms 
vanish for all measured observables we obtain stable fits for the coefficients $a_i$ that also 
obey \eqref{eq:constraint}. Moreover, we have experimented with different basis sets of 
integration cycles and found consistent results. As an additional test, we have also performed 
fit analyses based on sets of $N_\gamma<3$ integration cycles and find that 
the fit works as long as the integration cycles that are sampled in a simulation 
are linear combinations of the cycles used for the fit, which is the expected behavior. When choosing 
$N_\gamma>3$, on the other hand, fits are no longer reliable since the $\langle z^n\rangle_{\gamma_i}$
for different $i$ are not linearly independent in this case.

Overall, our results suggest that the question which linear combination of basis integration 
cycles is sampled in a complex Langevin simulation of \eqref{eq:quartic_1d} with $K=1$ and 
vanishing boundary terms has a unique answer, which can be found within the statistical 
uncertainties via an appropriate least-squares fit. In particular, we find the fit to be good 
(according to the criteria of \cref{app:chi_squared}) if and only if it produces this unique 
answer. Our results thus provide a sound test of the theorem \eqref{eq:salcedo_seiler}, which 
is found to hold in exactly the way one would expect.

\subsubsection{Nontrivial kernel}
We have seen that for most values of $\lambda$, measurements in complex Langevin simulations 
with $K=1$ are either affected by boundary terms (i.e., slowly decaying probability 
distributions) or by the sampling of unwanted integration cycles. However, as is known since  
\cite{OOS89}, the introduction of a nontrivial kernel $K\approx\lambda^{-1/2}$ in 
\eqref{eq:cle} can ensure 
$\langle z^n\rangle_{\mathrm{CL}}=\langle z^n\rangle_{\mathrm{exact}}$ for arbitrary 
$\lambda\neq0$. This is because -- in general -- a kernel has the effect of rotating the 
simulation trajectories in a certain way. In fact, as we shall see below, for the choice 
$K=\lambda^{-1/2}$ the ensuing probability distribution is confined to a one-dimensional 
subset of the complex plane, namely the line passing through the origin at an angle given by 
the argument of $\lambda^{-1/4}$. This implies that a simulation with this choice of kernel is 
equivalent to a real Langevin simulation in the variable $\xi$ introduced in 
\eqref{eq:variable_transformation}, which can be shown to converge. More generally, the 
argument of the kernel (or more precisely, the argument of the noise term $\sqrt{K}\eta$) can 
be expected to govern the orientation of the distribution of $z$ in the complex plane. This 
also means that the kernel, apart from being able to get rid of boundary terms, must also 
affect the coefficients $a_i$ in \eqref{eq:salcedo_seiler} in some way. To the best of our 
knowledge, the precise way in this happens was first studied in \cite{MHS24p}. Here, we shall 
extend these results. 

To this end, we focus on $l=5$, i.e., $\lambda=e^{5\ii\pi/6}$ in \eqref{eq:quartic_1d}, but 
mention that here and in the remainder of this work we may have as well chosen any other value 
of $\lambda\neq0$ with only minor modifications in the analysis. The effect of different 
kernels, i.e., different values of $m$ in \eqref{eq:kernel_1d}, on the distribution of $z$ can 
be appreciated from the histograms shown in \cref{fig:histograms_1d}. Indeed, we observe that 
the distributions are aligned roughly parallel to $\sqrt{K}$. Note, however, the different 
shapes of the distributions for different values of $m$. Two observations are particularly 
striking: 

\begin{figure}[t]
    \centering
    \includegraphics[width=\linewidth]{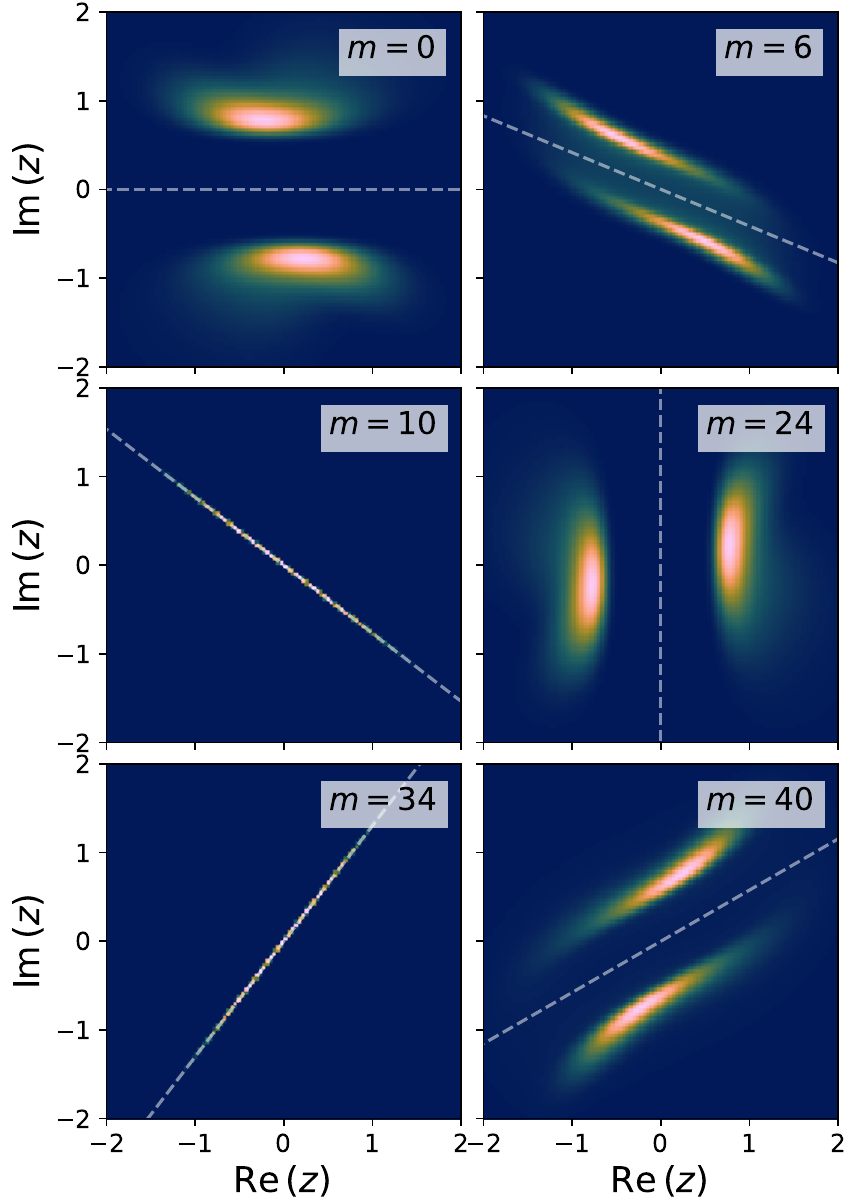}
    \caption{Histograms of $z$ in the complex plane, obtained in complex Langevin simulations 
             of the model \eqref{eq:quartic_1d} with $\lambda=e^{5\ii\pi/6}$ and different 
             kernels $K=e^{-m\ii\pi/24}$. The brighter regions indicate a higher probability 
             and the dashed lines lie in the direction of the argument of the noise 
             coefficient $\sqrt{K}$.}
    \label{fig:histograms_1d}
\end{figure}

First of all, the distribution is confined to a line not only for $m=10$, but for $m=34$ as 
well. Notice that in those cases all simulation trajectories converge to the same line 
(determined by $\sqrt{K}$) despite the use of random initial conditions. Under closer 
inspection, it turns out that the two lines sampled for $m=10$ and $m=34$ are precisely the 
two stable Lefschetz thimbles\footnote{By a Lefschetz thimble we refer to a one-dimensional 
subset of the complex plane going through a critical point of the action $S$ and along which 
$\im\,S=const.$. For the quartic model \eqref{eq:quartic_1d}, there are four such thimbles: 
two stable ones, on which $\re\,S$ increases as $z$ approaches complex infinity and two 
unstable ones, on which it decreases.} of the model \eqref{eq:quartic_1d} for this value of 
$\lambda$. Due to the way we have defined the real integration cycle $\gamma_1$, it is thus 
only natural to conclude that $m=10$ is the best choice for reproducing 
$\langle z^n\rangle_{\mathrm{exact}}$ in a simulation. In fact, it is a general expectation 
that if the complex Langevin distribution aligns with the relevant Lefschetz thimbles of 
the theory, the produced results will likely be correct. For previous studies on the relevance 
of Lefschetz thimbles within complex Langevin simulations, see, e.g., 
\cite{Aar13,ABS14,ALR23,BHM24}.

Second, for certain ranges of $m$, for instance around $m=0$ or $m=24$, the distributions 
consist of two disconnected parts separated by a barrier, with a zero transition probability 
between them in extreme cases. This can happen because the drift vanishes or points away from 
the barrier in its vicinity, while the noise is perpendicular to it. It results in the loss of 
ergodicity, implying that a single simulation would be stuck in one of the two regions, 
depending on the initial conditions, forever, spoiling expectation values. The reason why we 
nevertheless observe symmetric distributions is that we take the histograms over all our 
parallel simulations, each of which has randomized initial conditions, in many successive 
runs. 

The latter observation has important consequences for the interpretation of the computed 
coefficients $a_i$, i.e., for the question which integration cycles are sampled in our 
simulation runs. As a matter of fact, we have verified that one may obtain different results 
for the $a_i$ if one uses the same fixed (rather than different randomized) initial 
conditions for all simulations, such that only one of the two high-probability regions 
contributes. In particular, in that case one may find $a_3\neq0$, indicating that individual 
simulations, which are affected by the ergodicity problem, may actually sample $\gamma_3$, 
even though its contributions cancel in the average over many simulations when randomized 
initial conditions are used. Indeed, we find $a_3$ to be vanishingly small in all our 
simulations with random initial conditions, see, e.g., \cref{tab:coefficients_1d}. The reason 
for this likely has to do with symmetry considerations, as with random initial conditions we 
find all distributions to be symmetric under the reflection $z\to-z$, under which $\gamma_1$ 
and $\gamma_2$ are invariant (up to inversion), but $\gamma_3$ changes in a topologically 
nontrivial way, turning into a different integration cycle.

Generally speaking, there is a certain correlation between the shape of the distribution and 
the integration cycles that contribute to expectation values. For instance, as we have 
mentioned, the line sampled for $m=10$ is precisely the integration contour used to compute 
$\langle z^n\rangle_{\gamma_1}$. This statement should not be taken at face value, however. 
Indeed, a distribution that is close (or parallel) to a straight line that lies within the 
equivalence class of paths described by a certain integration cycle $\gamma_i$ does not 
automatically imply that $\gamma_i$ is the only cycle that contributes. However, it does 
provide an explanation for the fact that simulations with $K=1$, for which the distributions 
are approximately parallel to the real line, are plagued by nonzero boundary terms if 
$\re\,\lambda<0$. After all, for this range of $\lambda$ the real line is not homotopic 
to any valid integration cycle.

\begin{figure}[t]
    \centering
    \includegraphics[width=\linewidth]{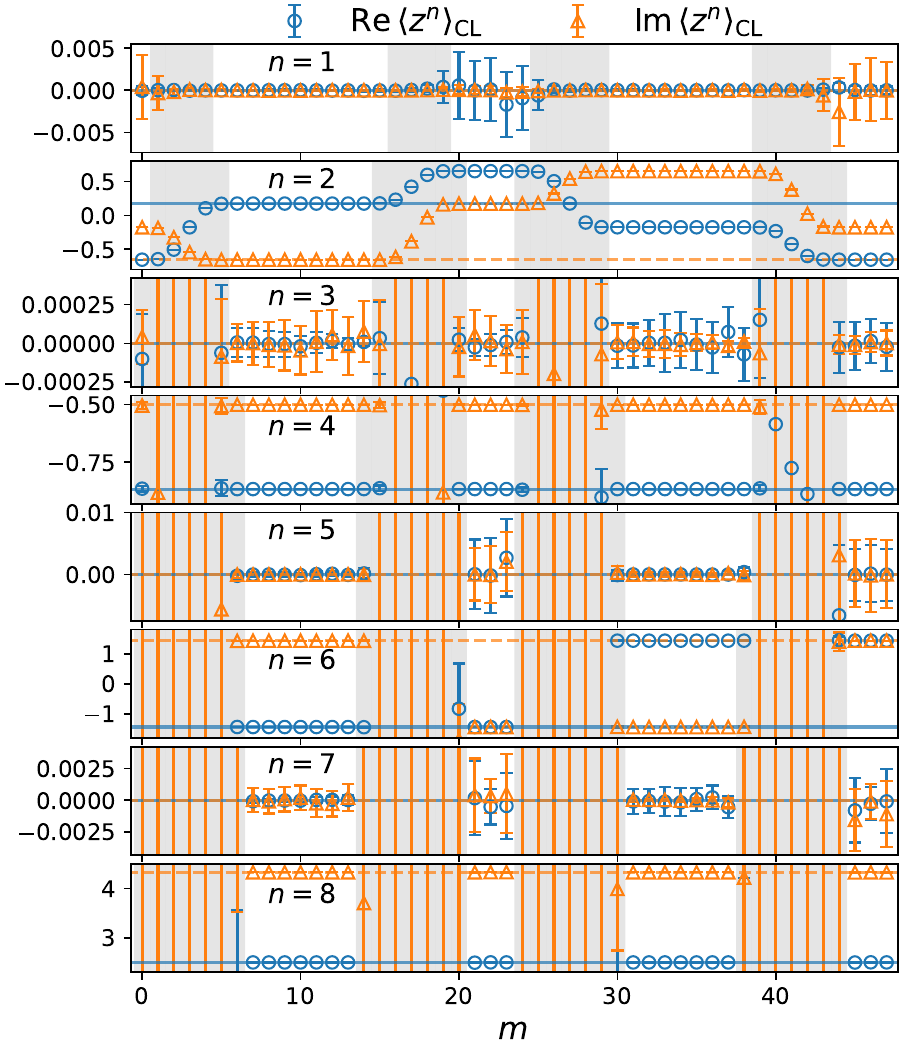}
    \caption{Real and imaginary parts of the observables $\langle z^n\rangle_{\mathrm{CL}}$ in 
             the theory \eqref{eq:quartic_1d} with $l=5$ in \eqref{eq:lambda} for different 
             $n$ as a function of the kernel parameter $m$ in \eqref{eq:kernel_1d}. For values 
             of $m$ within the shaded regions, the corresponding observable has a nonzero 
             boundary term. The solid and dashed horizontal lines indicate the real and 
             imaginary parts of the corresponding exact results \eqref{eq:exact_1d}, 
             respectively.}
    \label{fig:observables_vs_m_1d}
\end{figure}

In \cref{fig:observables_vs_m_1d}, we show the observables $\langle z^n\rangle_{\mathrm{CL}}$ 
as a function of the kernel parameter $m$. Once again, values of $m$ for which an observable 
has a nonzero boundary term are indicated by shaded regions. Many features of 
\cref{fig:observables_vs_m_1d} are similar to 
\cref{fig:observables_vs_l_1d_even,fig:observables_vs_l_1d_odd}, such as the signals for 
nonvanishing boundary terms becoming clearer (and the presence of boundary terms being 
correlated with larger uncertainties) for larger values of $n$.

The most striking feature of \cref{fig:observables_vs_m_1d}, however, was already observed in 
\cite{OOS89}. Namely, as was mentioned before, while there are boundary terms and thus 
simulation results inconsistent with $\langle z^n\rangle_{\mathrm{exact}}$ for the trivial 
kernel $K=1$ ($m=0$), a nontrivial kernel can actually ensure that 
$\langle z^n\rangle_\CL=\langle z^n\rangle_{\mathrm{exact}}$. This is true especially in the 
vicinity of $K=\lambda^{-1/2}$ ($m=2l$), as anticipated. More precisely, there is a large 
plateau, of roughly the size $7$, around $m=10$, for which the simulation results are 
compatible with \eqref{eq:exact_1d}. Moreover, there is another such plateau around $m=34$, 
which plays a similarly special role (see the discussion of \cref{fig:histograms_1d}), except 
that it is not associated with $\langle z^n\rangle_{\mathrm{exact}}$, and for which boundary 
terms vanish as well. Finally, there are two more, albeit smaller, plateaus, around $m=22$ and 
$m=46$, respectively. Here, too, boundary terms are consistent with zero but 
$\langle z^n\rangle_\CL\neq\langle z^n\rangle_{\mathrm{exact}}$ for certain $n$. Indeed, as 
before, for $n$ either being odd or a multiple of four, the complex Langevin results actually 
agree with \eqref{eq:exact_1d} on all plateaus due to the symmetry of those observables. Thus, 
when we say that simulation results are consistent with \eqref{eq:exact_1d}, what we mean is 
that $\langle z^n\rangle_\CL=\langle z^n\rangle_{\mathrm{exact}}$ for all $n$.

We believe that the fact that there are extended regions in $m$ (and thus in the kernel $K$) 
for which boundary terms vanish is a nontrivial finding. Perhaps even more importantly, the 
observation that simulation results are consistent with the desired values 
\eqref{eq:exact_1d} on a large plateau in the kernel parameter could prove valuable for future 
investigations of more realistic systems using complex Langevin simulations with a nontrivial 
kernel.

\begin{figure}[t]
    \centering
    \includegraphics[width=\linewidth]{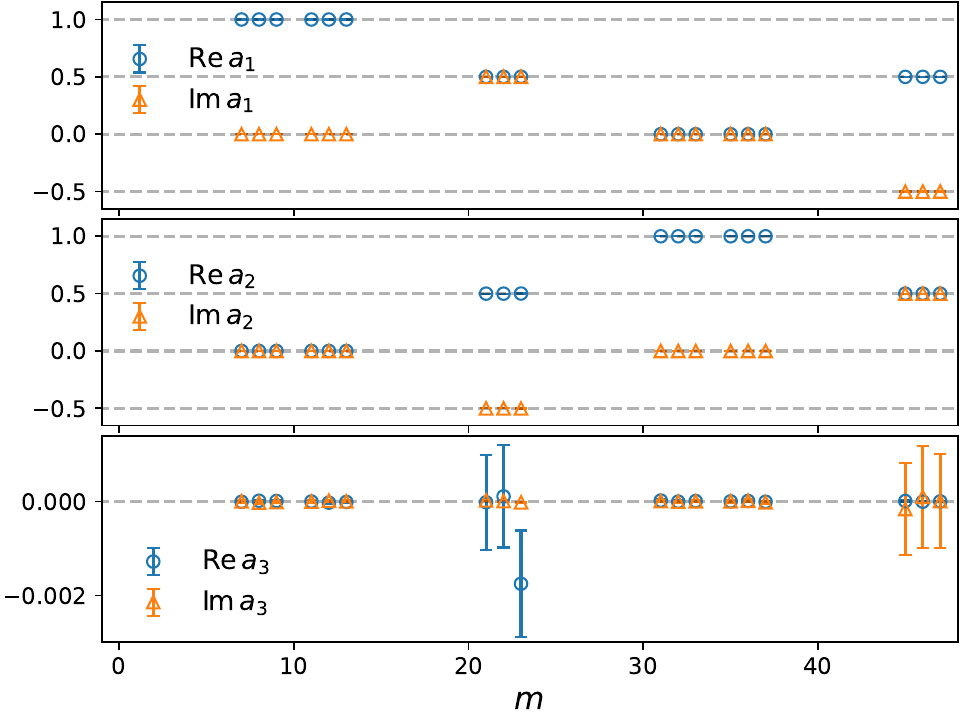}
    \caption{Real and imaginary parts of the coefficients $a_i$ in the model          
             \eqref{eq:quartic_1d} with $\lambda=e^{5\ii\pi/6}$ as a function of the kernel 
             parameter $m$ in \eqref{eq:kernel_1d}, obtained via the least squares fit 
             procedure outlined in \cref{sec:measurement_coefficients} using the observables 
             \eqref{eq:observable_set_1d}. We plot only those values of $m$ for which boundary 
             terms are consistent with zero. The points $m=10$ and $m=34$, for which trivially 
             $a_i=\delta_{i1}$ and $a_i=\delta_{i2}$, respectively, are not shown either, as 
             the corresponding coefficients cannot be computed with our approach. The dashed 
             horizontal lines are placed in steps of $1/2$ to guide the eye.}
    \label{fig:coefficients_vs_m_1d}
\end{figure}

The findings presented so far lead one to expect that the coefficients $a_i$ in 
\eqref{eq:salcedo_seiler} should read $a_i\approx\delta_{i1}$ in the vicinity of $m=10$ but 
take different values on the other plateaus. To substantiate this claim, we plot the $a_i$ as 
a function of $m$ in \cref{fig:coefficients_vs_m_1d}. Only points for which a reasonable fit 
could be obtained are shown. As before, these points correspond to the ones for which boundary 
terms are consistent with zero.
We observe that there are three different scenarios, depending on the kernel parameter $m$: 
First, around $m=10$, the real integration cycle $\gamma_1$ dominates and contributions from 
$\gamma_{i\neq1}$ are clearly subleading, as expected from \cref{fig:histograms_1d}. Second, 
close to $m=34$, it is $\gamma_2$, i.e., the imaginary integration cycle, that (almost) 
exclusively contributes to the complex Langevin results. This, again, can be appreciated from 
\cref{fig:histograms_1d}. Finally, on the smaller plateaus around $m=22$ and $m=46$, 
respectively, the simulations sample two different linear combinations of $\gamma_1$ and 
$\gamma_2$, which, however, is less obvious from \cref{fig:histograms_1d}. As before, we find 
the condition \eqref{eq:constraint} to be true and $a_3$ to be consistent with zero in all of 
these scenarios. The reason why both the real and imaginary parts of all coefficients 
seem to only take values that are multiples of $1/2$ is not clear to us at this point.

We conclude that the theorem \eqref{eq:salcedo_seiler} holds in all cases we have investigated 
so far. While straightforward complex Langevin simulations with $K=1$ often suffer from 
wrong convergence, this can be cured via an appropriate choice of kernel. In particular, we 
have seen that a kernel of the form \eqref{eq:kernel_1d} can be used to, on the one hand, 
remove boundary terms and, on the other hand, affect which integration cycles contribute to 
$\langle z^n\rangle_\CL$. Hence, in a sense, a kernel can be used to generate rotations in 
the space of integration cycles. The reason that a kernel as simple as \eqref{eq:kernel_1d} is 
sufficient to restore correct convergence is owed to the simplicity of the model 
\eqref{eq:quartic_1d}. In more complex models, this might no longer be true and more 
sophisticated, i.e., $z$-dependent kernels may have to be used. This has been done 
successfully, e.g., in \cite{OSZ91}, and we are planning to re-investigate this issue from the 
point of view of integration cycles in future work. We remark, however, that the existence of 
a kernel that can ensure correct convergence is -- to the best of our knowledge -- not 
guaranteed in general theories.

Up to now, we have mainly provided numerical confirmation for the theorem 
\eqref{eq:salcedo_seiler}, which was anyways proven to be true in one dimension in 
\cite{SS19}. In the next subsection, however, we test its validity when two degrees of freedom 
are involved. As far as we know, no corresponding proof exists in this case, such that our 
numerical results are the first hints towards a generalization of \eqref{eq:salcedo_seiler} 
to arbitrary dimensions, i.e., arbitrary theories.

\subsection{Two dimensions}\label{sec:results_2d}
We consider the following class of two-dimensional models:
\begin{equation}\label{eq:quartic_2d}
    S_a(z_1,z_2) = \frac{\lambda}{4}\left(z_1^4+z_2^4+az_1^2z_2^2\right)\;,
\end{equation}
where $z_i\in\C$, $\lambda$ is defined as in \eqref{eq:lambda} and $a\in\R$ is a tunable 
parameter controlling the amount of coupling between $z_1$ and $z_2$. In the following, we 
shall be interested in observables of the form $\langle z_1^{n_1}z_2^{n_2}\rangle$, where the 
$n_i$ are non-negative integers. As compared to the one-dimensional case, however, for these 
observables (and $a\neq0$) there exist no analytical results analogous to \eqref{eq:exact_1d}, 
such that we have to resort to numerical integration. In the simulations discussed below, we 
have measured $\langle z_1^{n_1}z_2^{n_2}\rangle$ for those combinations of $n_1$ and $n_2$ 
with $n_1+n_2\leq4$. We shall refer to observables as even if both $n_1$ and $n_2$ are even 
numbers and as odd otherwise. We note that the boundary term expectation values in two 
dimensions are defined as in \eqref{eq:boundary_terms}, with the difference that $L_c$ is now 
defined as a sum over degrees of freedom, i.e., 
$L_c=\sum_{i=1}^2\left(\partial_i-\partial_iS\right)K_i\partial_i$.

We find it adequate at this point to discuss to which extent the two models 
\eqref{eq:quartic_1d} and \eqref{eq:quartic_2d} can actually be compared with one another. 
After all, these theories have different numbers of degrees of freedom as well as symmetries. 
Nevertheless, both actions are polynomials of quartic order with no terms of lower or higher 
order involved, suggesting that they might at the very least share some common features. 
Needless to say, this question is highly dependent on the coupling parameter $a$: 

On the one hand, for $a=0$, the resulting free theory $S_0$ decomposes into two independent 
one-dimensional models of the form \eqref{eq:quartic_1d}. For complex Langevin simulations 
with a diagonal kernel, as in our setup, the lack of coupling between $z_1$ and $z_2$ means 
that their respective evolutions are completely decoupled from one another, such that their 
probability distributions are simply the one-dimensional ones we have already studied (c.f.
\cref{fig:histograms_1d}). This decoupling also makes the discussion of integration cycles 
particularly simple. Indeed, for $a=0$, one may write the two-dimensional basis integration 
cycles in terms of the one-dimensional ones as $\gamma_i\times\gamma_j$, where the 
notation indicates that one may integrate over $z_1$ and $z_2$ (along $\gamma_i$ and 
$\gamma_j$, respectively) independently, since the integrals factorize. More precisely, one 
finds
\begin{equation}\label{eq:factorization}
    \langle z_1^{n_1}z_2^{n_2}\rangle_{\gamma_i\times\gamma_j} = 
        \langle z^{n_1}\rangle_{\gamma_i}\cdot\langle z^{n_2}\rangle_{\gamma_j}\;,
\end{equation}
where each factor on the right-hand side is a one-dimensional integral of the form 
\eqref{eq:cycle_integral_1d}. Hence, there are $N_\gamma=3\cdot3=9$ independent integration 
cycles in the theory $S_0$. This also means that the entire analysis of \cref{sec:results_1d} 
carries over directly. For more details on this subject, we refer to 
\cref{app:counting_cycles}. 

As one varies $a$ and, thus, introduces a nonzero coupling between $z_1$ and $z_2$, however, 
these conclusions might no longer hold. Indeed, for large values of $a$, the asymptotics of 
\eqref{eq:quartic_2d}, which determine the possible integration cycles, are no longer governed 
by the $z_i^4$ but by the mixing term $z_1^2z_2^2$ instead. As is discussed in 
\cref{app:counting_cycles}, the dependence of the number of independent integration cycles 
$N_\gamma$ on $a$ has the following peculiar behavior in the model \eqref{eq:quartic_2d}:
\begin{equation}\label{eq:number_of_cycles}
    N_\gamma = 
        \begin{cases}
            9 \quad \mathrm{if} \quad \vert a\vert < 2\;,\\
            2 \quad \mathrm{if} \quad \vert a\vert \geq 2\;.
    \end{cases}
\end{equation}
The two points $a=\pm2$ are thus special in that they mark the transition from a weak 
coupling regime, in which the two-dimensional theory bears some resemblance to two 
independent one-dimensional theories (becoming an equivalence for $a=0$), to a strong 
coupling regime in which the interaction term dominates. It is worth emphasizing that the 
strong-coupling regime is actually simpler than the free theory in terms of the number of 
relevant integration cycles. We shall now investigate different scenarios, depending on the 
value of $a$, in more detail.

\subsubsection{\texorpdfstring{$\mathrm{O}(2)$}{O(2)}-symmetric case}
We start by discussing the case $a=2$, i.e., the model 
\begin{equation}\label{eq:quartic_2d_o2}
    S_2(z_1,z_2) = \frac{\lambda}{4}(z_1^2+z_2^2)^2\;,
\end{equation}
which features an $\mathrm{O}(2)$ symmetry. In order to define the two linearly independent 
integration cycles for this model, we introduce two variables $\xi_1$ and $\xi_2$ in analogy 
to \eqref{eq:variable_transformation}. Then, as before, we define $\gamma_1$ to be the real 
integration cycle in these new variables. More precisely, $\gamma_1$ is the equivalence class 
of integration surfaces that are homotopic to the two-dimensional real submanifold of 
$\C^2\ni(\xi_1,\xi_2)$. Similarly, we define $\gamma_2$ to be the corresponding imaginary 
cycle. In the notation of \cref{eq:factorization}, these two basis cycles may be represented 
as $\gamma_1\times\gamma_1$ and $\gamma_2\times\gamma_2$, respectively, where we use the 
same symbol for the one- and two-dimensional cycles. As before, we define 
$\langle z_1^{n_1}z_2^{n_2}\rangle_{\mathrm{exact}}:=\langle z_1^{n_1}z_2^{n_2}\rangle_{\gamma_1}$, 
but emphasize once again that this is merely a choice; the result of the 
$\gamma_1$-integration is by no means more correct than the one over $\gamma_2$. 

\begin{figure}[t]
    \centering
    \includegraphics[width=1\linewidth]{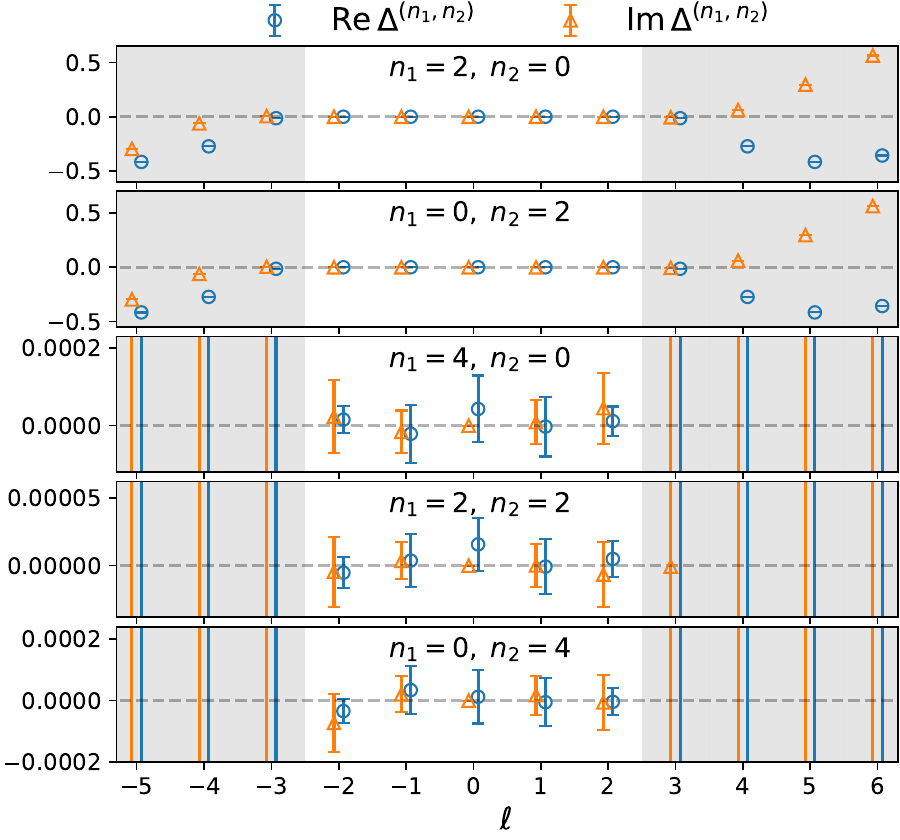}
    \caption{Similar as in \cref{fig:observables_vs_l_1d_even} but for $\Delta^{(n_1,n_2)}$, 
             as defined in \eqref{eq:delta_2d}, and in the model \eqref{eq:quartic_2d_o2}.}
    \label{fig:observables_vs_l_2d_even}
\end{figure}

Given the success of our analysis in the one-dimensional model \eqref{eq:quartic_1d}, we now 
essentially repeat the steps of \cref{sec:results_1d} for the theory \eqref{eq:quartic_2d_o2}. 
To begin with, we employ a trivial kernel, $K_1=K_2=1$, in \eqref{eq:discrete_cle}. Moreover, 
we once again introduce the difference
\begin{equation}\label{eq:delta_2d}
    \Delta^{(n_1,n_2)} := 
    \langle z_1^{n_1}z_2^{n_2}\rangle_\CL - \langle z_1^{n_1}z_2^{n_2}\rangle_{\mathrm{exact}}
\end{equation}
to quantify the deviations between complex Langevin and exact results.
As a two-dimensional analog of \cref{fig:observables_vs_l_1d_even}, we plot 
$\Delta^{(n_1,n_2)}$ as a function of the parameter $l$ in \eqref{eq:lambda} for the even 
observables in \cref{fig:observables_vs_l_2d_even}. Most of the conclusions drawn in the 
context of \cref{fig:observables_vs_l_1d_even} also hold true in the present case, such that 
we refrain from repeating them. Note, however, that, while there were no boundary terms for 
$l=6$ in the one-dimensional case, this is no longer true in the model 
\eqref{eq:quartic_2d_o2}. This has to do with the way two-dimensional integration cycles are 
related to their one-dimensional counterparts and shall be discussed in more detail below. As 
a consequence, we find agreement between simulation and exact results for all values of 
$\lambda$ without boundary terms. We do not show the odd observables here, as we find them all 
to be consistent with zero whenever boundary terms vanish. Under the assumption that 
\eqref{eq:salcedo_seiler} indeed holds in two dimensions, this is not surprising, given that 
$\langle z_1^{n_1}z_2^{n_2}\rangle_{\gamma_1}=\langle z_1^{n_1}z_2^{n_2}\rangle_{\gamma_2}=0$ 
if at least one of $n_1$ and $n_2$ is odd. 

\begin{figure}[t]
    \centering
    \includegraphics[width=1\linewidth]{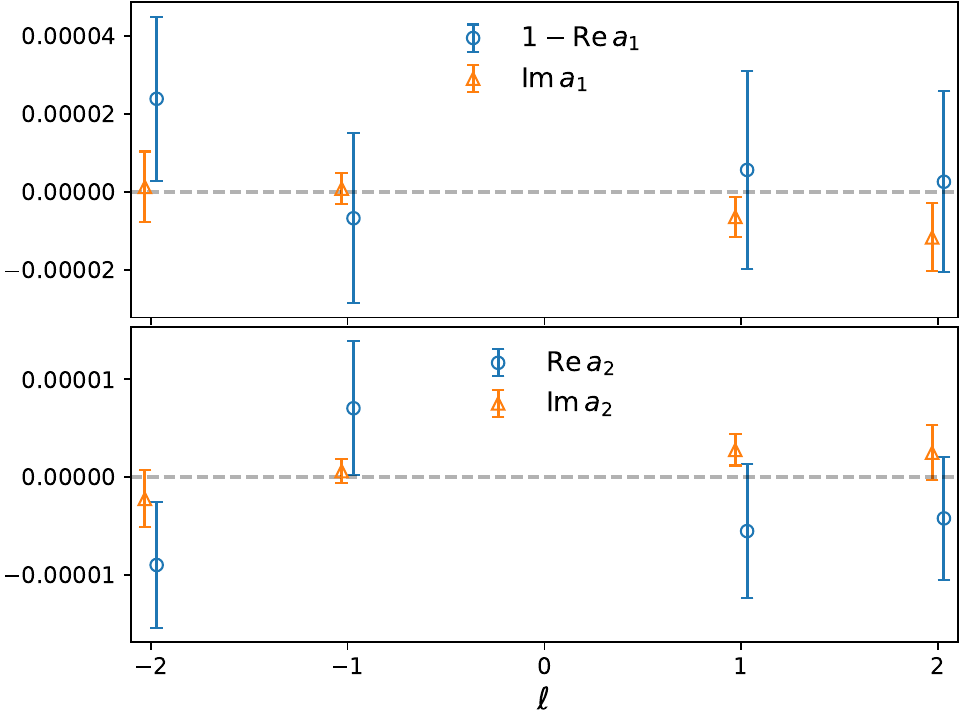}
    \caption{Real and imaginary parts of the coefficients $a_i$ in the model          
             \eqref{eq:quartic_2d_o2} as a function of the  
             parameter $l$ in \eqref{eq:lambda}, obtained via the least-squares fit procedure 
             outlined in \cref{sec:measurement_coefficients} using the observables 
             \eqref{eq:observable_set_2d_even}. We only plot those values of $l\neq0$ for 
             which boundary terms are consistent with zero. The point $l=0$, for which 
             $a_i=\delta_{i1}$, trivially, is not shown either as the corresponding 
             coefficients cannot be computed with our approach. The dashed horizontal lines 
             indicate zero. Note that the upper panel shows $1-\re\,a_1$ and the data points 
             have been displaced horizontally, both for reasons of 
             readability.}
    \label{fig:coefficients_vs_l_2d}
\end{figure}

The latter fact also implies that we have to restrict to the even observables for the 
computation of the coefficients $a_i$ in \eqref{eq:salcedo_seiler}. Thus, we choose the 
following set of observables, being the largest one possible, for the least-squares fit:
\begin{equation}\label{eq:observable_set_2d_even}
    \left\{\langle z_1^2\rangle, \langle z_2^2\rangle, \langle z_1^4\rangle, 
    \langle z_1^2z_2^2\rangle, \langle z_2^4\rangle
    \right\}\;.
\end{equation}
With this and $N_\gamma=2$, we proceed as before in order to compute the $a_i$ and we show 
their $l$-dependence in \cref{fig:coefficients_vs_l_2d}. We find $a_1\approx1$ and 
$a_2\approx0$ for all $\lambda$ with vanishing boundary terms, while the fits become unstable 
when boundary terms are nonzero. This is a nontrivial finding and a first hint of the validity 
of \eqref{eq:salcedo_seiler} beyond one dimension. It is also noteworthy that we do not 
observe any nontrivial linear combinations of $\gamma_1$ and $\gamma_2$ here, even though they 
would be allowed according to \eqref{eq:salcedo_seiler}. While, as we have mentioned, one may 
not easily establish a direct connection between the models \eqref{eq:quartic_1d} and 
\eqref{eq:quartic_2d_o2}, we nonetheless conclude that one may encounter situations where the 
analysis of integration cycles simplifies rather dramatically when increasing the number of 
degrees of freedom.

As in one dimension, complex Langevin simulations of \eqref{eq:quartic_2d_o2} with a trivial 
kernel cannot reproduce the exact results for most values of $\lambda$, in particular those 
with $\re\,\lambda<0$, due to the presence of boundary terms. In an attempt to nonetheless 
obtain an answer for these values, we thus once again introduce a nontrivial kernel 
and consider $l=5$ in \eqref{eq:lambda} as a representative. While, in principle, we may 
choose $K_1\neq K_2$ in \eqref{eq:discrete_cle}, we instead take both kernel elements to be equal for 
the time being. This implies that $z_1$ and $z_2$ are being treated on completely equal 
footing, which must be reflected in their respective distributions coinciding. Moreover, we parametrize 
$K:=K_1=K_2$ as in \eqref{eq:kernel_1d}, using an integer $m$.

\begin{figure}[t]
    \centering
    \includegraphics[width=\linewidth]{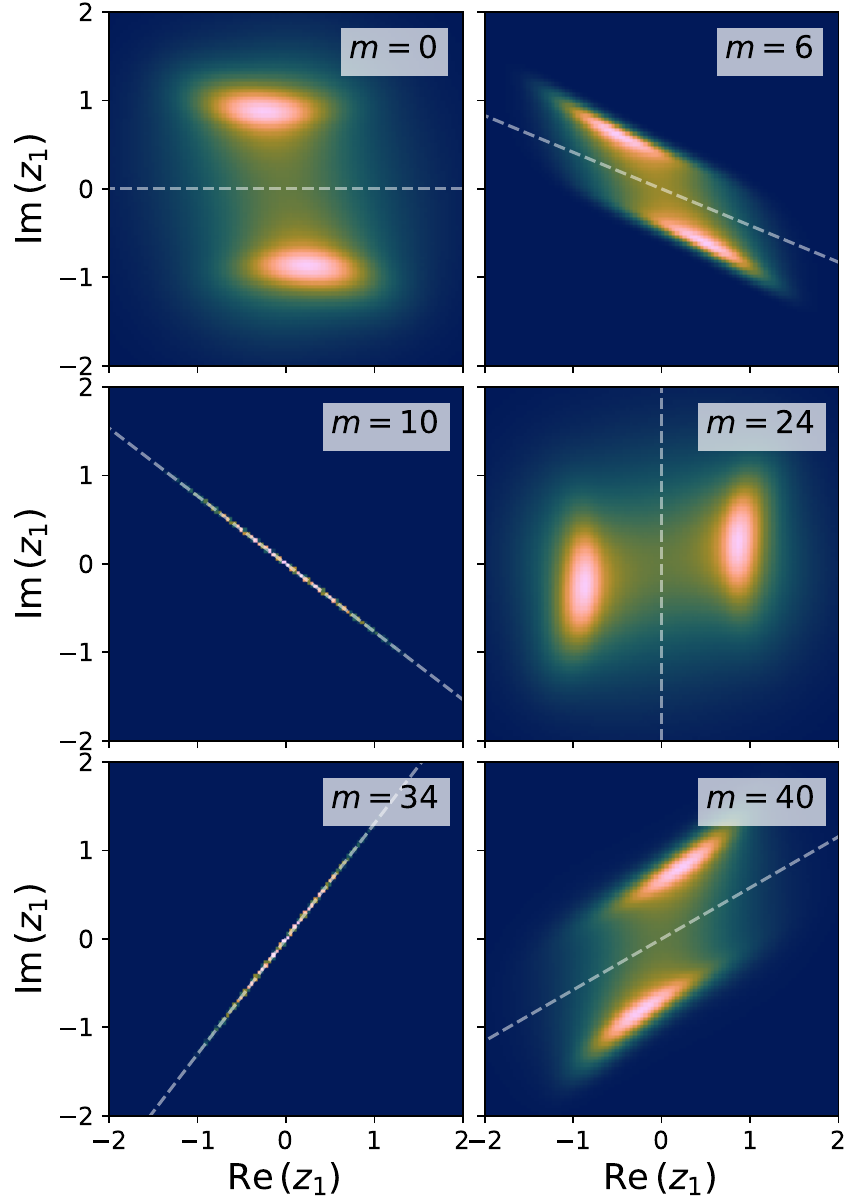}
    \caption{Histograms of $z_1$  obtained in complex Langevin simulations of the model  
             \eqref{eq:quartic_2d_o2} with $\lambda=e^{5\ii\pi/6}$ and different kernels 
             $K=e^{-m\ii\pi/24}$, chosen equal for both variables. The brighter regions 
             indicate a higher probability and the dashed lines lie in the direction of the 
             argument of the noise coefficients $\sqrt{K}$. The respective histograms of $z_2$ 
             are identical and thus not shown.}
    \label{fig:histograms_2d_equal_kernels}
\end{figure}

The effect of this choice of kernel on the probability distributions is shown in 
\cref{fig:histograms_2d_equal_kernels}, which is an analog of \cref{fig:histograms_1d} for the 
two-dimensional model \eqref{eq:quartic_2d_o2}. Since we have verified that the distributions 
are indeed identical for $z_1$ and $z_2$, we only show those of the former. The general 
features of the two-dimensional distributions closely resemble those discussed in 
\cref{fig:histograms_1d}. In particular, for $m=10$ and $m=34$, the distributions are again 
confined to lines, such that we expect values of $m$ close to $10$ to give the best agreement 
with $\langle z_1^{n_1}z_2^{n_2}\rangle_{\mathrm{exact}}$ due to our definition of $\gamma_1$. 
There is one major difference between 
\cref{fig:histograms_1d,fig:histograms_2d_equal_kernels}, however. Namely, in the latter we do 
not observe an ergodicity problem anymore, as the two high-probability regions are now 
connected to each other. This means that the strong interaction between $z_1$ and $z_2$ allows 
each of them to transition between the high-probability regions, which would not be possible 
for $a=0$. 

\begin{figure}[t]
    \centering
    \includegraphics[width=1\linewidth]{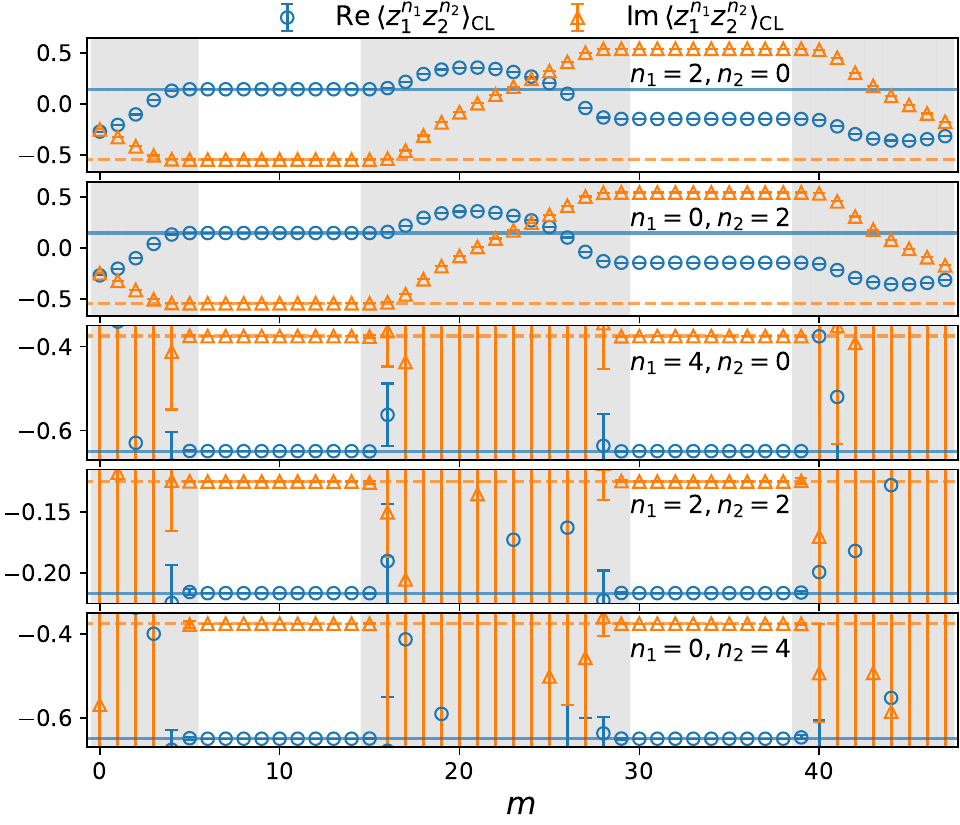}
    \caption{As in \cref{fig:observables_vs_m_1d} but for the observables 
             $\langle z_1^{n_1}z_2^{n_2}\rangle_{\mathrm{CL}}$, with $n_1$ and $n_2$ both 
             even, in the theory \eqref{eq:quartic_2d_o2}, using $K_1=K_2$, both parametrized 
             in terms of the integer $m$ in \eqref{eq:kernel_1d}.}
    \label{fig:observables_vs_m_2d_even}
\end{figure}

The $m$-dependence of the even observables is shown in \cref{fig:observables_vs_m_2d_even}. As 
before, we once again observe large plateaus in the vicinity of $m=10$ and $m=34$, 
respectively. In fact, these plateaus appear to be slightly larger than those in \cref{fig:observables_vs_m_1d}.
However, contrary to the one-dimensional case, the additional plateaus near 
$m=22$ and $m=46$ are absent here and instead one finds boundary terms for those values. We 
interpret this finding as follows: as we have mentioned, in one dimension, a single simulation 
with $m$ close to $22$ or $46$ may sample $\gamma_3$ due to the ergodicity problem, even 
though its contributions cancel in the average over simulations. Something similar would 
happen in the model \eqref{eq:quartic_2d} for small values of $a$, for which appropriate 
extensions of $\gamma_3$ to two dimensions still constitute valid integration cycles. For 
$a=2$, however, this is no longer true, implying that such a 
choice of kernel forces the system onto a kind of unstable trajectory, resulting in nonzero 
boundary terms. This is also the reason why there are boundary terms for $l=6$ in 
\cref{fig:observables_vs_l_2d_even}, while they are absent in 
\cref{fig:observables_vs_l_1d_even,fig:observables_vs_l_1d_odd}. Once again, we refrain from 
plotting the odd observables, as they are trivial.

\begin{figure}[t]
    \centering
    \includegraphics[width=1\linewidth]{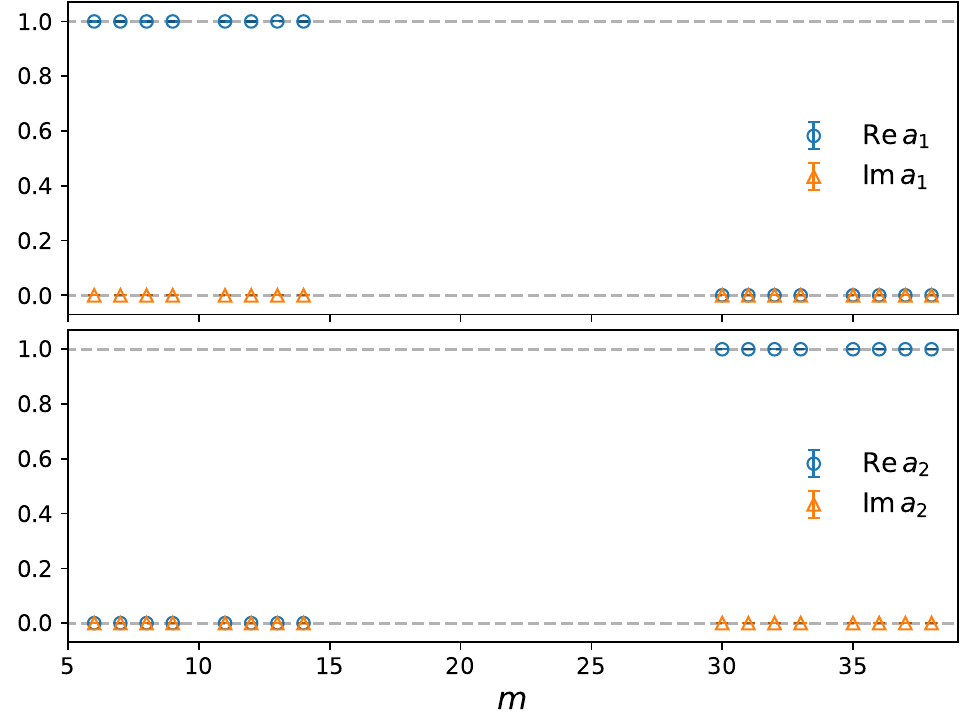}
    \caption{Analogous to \cref{fig:coefficients_vs_m_1d} but for the model \eqref{eq:quartic_2d_o2} with 
             $\lambda=e^{5\ii\pi/6}$ and using $K_1=K_2$, each parametrized by the integer $m$ in 
             \eqref{eq:kernel_1d}. For the least-squares fit we have used the observables 
             \eqref{eq:observable_set_2d_even} and $N_\gamma=2$. The dashed horizontal lines at $0$ and $1$ 
             are there to guide the eye.}
    \label{fig:coefficients_vs_m_2d}
\end{figure}

In \cref{fig:coefficients_vs_m_2d}, we show fits for the two coefficients $a_1$ and $a_2$ as a function of 
$m$. As expected from the previous discussion, we find that $\gamma_1$ dominates on the plateau around 
$m=10$, while $\gamma_2$ becomes dominant in the vicinity of $m=34$. As for $K_i=1$, we once again do not 
observe any nontrivial linear combinations involving both $\gamma_1$ and $\gamma_2$. Moreover, as in all 
previous cases, we obtain stable fits if and only if boundary terms are absent, which makes another strong 
point for the validity of \eqref{eq:salcedo_seiler} in general theories. Once again, the dramatic effect 
that different choices of kernel can have on the coefficients $a_i$ can hardly be overemphasized.

\subsubsection{General couplings}
We have seen before that the class of models \eqref{eq:quartic_2d} behaves in a qualitatively different way 
depending on whether $a$ is smaller or larger than two, since the number of independent integration cycles 
jumps from $9$ to $2$ at that value according to \eqref{eq:number_of_cycles}. As the final analysis of this 
work, we shall devote this subsection to a discussion of how this transition is reflected on the level of 
complex Langevin simulations by studying distributions of $z_1$ and $z_2$ for different values of $a$. 

In the following, we once again consider $\lambda=e^{5\ii\pi/6}$ in \eqref{eq:quartic_2d}, i.e., $l=5$ in 
\eqref{eq:lambda}. Now, however, we allow the two kernel elements $K_1$ and $K_2$ to vary independently of each 
other. To this end, we introduce two integers $m_1$ and $m_2$, parametrizing $K_1$ and $K_2$ as in 
\eqref{eq:kernel_1d}, respectively, but restrict them to the values $m_i\in\{10, 22, 34\}$. For $a<2$, this 
should allow us to probe a large variety of different integration cycles. In particular, for $a=0$ we have 
seen that the two-dimensional integration cycles trivially result from combinations of two independent one-
dimensional ones. Since in one dimension we have seen that we can generate contributions from $\gamma_1$ and 
$\gamma_2$ rather easily, we expect to be able to sample at least four different two-dimensional cycles for 
small values of $a$. We note that it is not our primary concern in this investigation to obtain simulation 
results that match 
$\langle z_1^{n_1}z_2^{n_2}\rangle_{\mathrm{exact}}:=\langle z_1^{n_1}z_2^{n_2}\rangle_{\gamma_1}$. After 
all, we know that this can be achieved with the choice $K_1=K_2=\lambda^{-1/2}$. Rather, we aim at obtaining 
a better understanding of how exactly the simulation trajectories are affected by nontrivial kernels and how 
this may influence the sampling of different integration cycles. 

\begin{figure}[t]
    \centering
    \includegraphics[width=\linewidth]{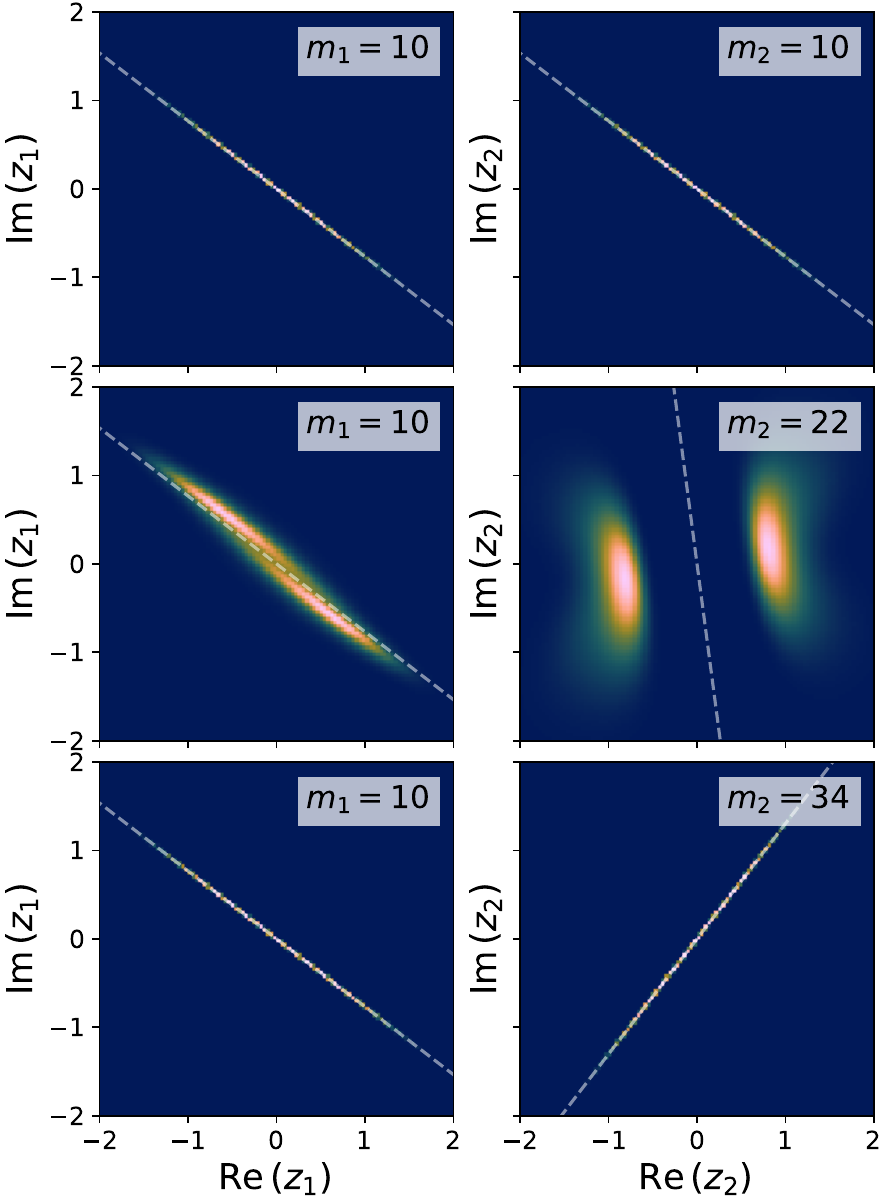}
    \caption{Histograms of $z_1$ (left column) and $z_2$ (right column) obtained in complex Langevin 
             simulations of the model \eqref{eq:quartic_2d} with $\lambda=e^{5\ii\pi/6}$ and $a=1$. The 
             kernel elements $K_1$ and $K_2$ are parametrized as in \eqref{eq:kernel_1d} using integers 
             $m_1$ and $m_2$, respectively. The top, center and bottom rows correspond to 
             $(m_1,m_2)=(10,10)$, $(10,22)$, and $(10,34)$, respectively. The brighter regions indicate a 
             higher probability and the dashed lines lie in the directions of the arguments of the 
             respective noise coefficients $\sqrt{K_i}$.}
    \label{fig:histograms_2d_unequal_kernels_a_1}
\end{figure}

Our choice of parametrizing $K_1$ and $K_2$ individually implies that the respective distributions of $z_1$ 
and $z_2$ need no longer be equivalent in general. A trivial example is given by $a=0$, for which the 
respective distributions of $z_1$ and $z_2$ are simply those shown in \cref{fig:histograms_1d} (with $m$ 
replaced by $m_1$ and $m_2$, respectively). Introducing a small coupling $a\neq0$ should then cause these 
distributions to wash out to a certain degree. This effect is seen in the center panel of 
\cref{fig:histograms_2d_unequal_kernels_a_1}, in which we show histograms of $z_1$ and $z_2$ that result 
from different combinations of $m_1$ and $m_2$ at an intermediate interaction strength $a=1$. Observe, 
however, that no such distortion occurs if both $m_i\in\{10,34\}$. For these particular choices of kernel, 
both distributions remain confined to their respective lines despite the interaction. We emphasize that this 
does not imply that the trajectories of $z_1$ and $z_2$ are independent of one another. On the contrary, 
their correlations increase with $a$ irrespectively of the one-dimensional confinement. 

\begin{figure}[t]
    \centering
    \includegraphics[width=\linewidth]{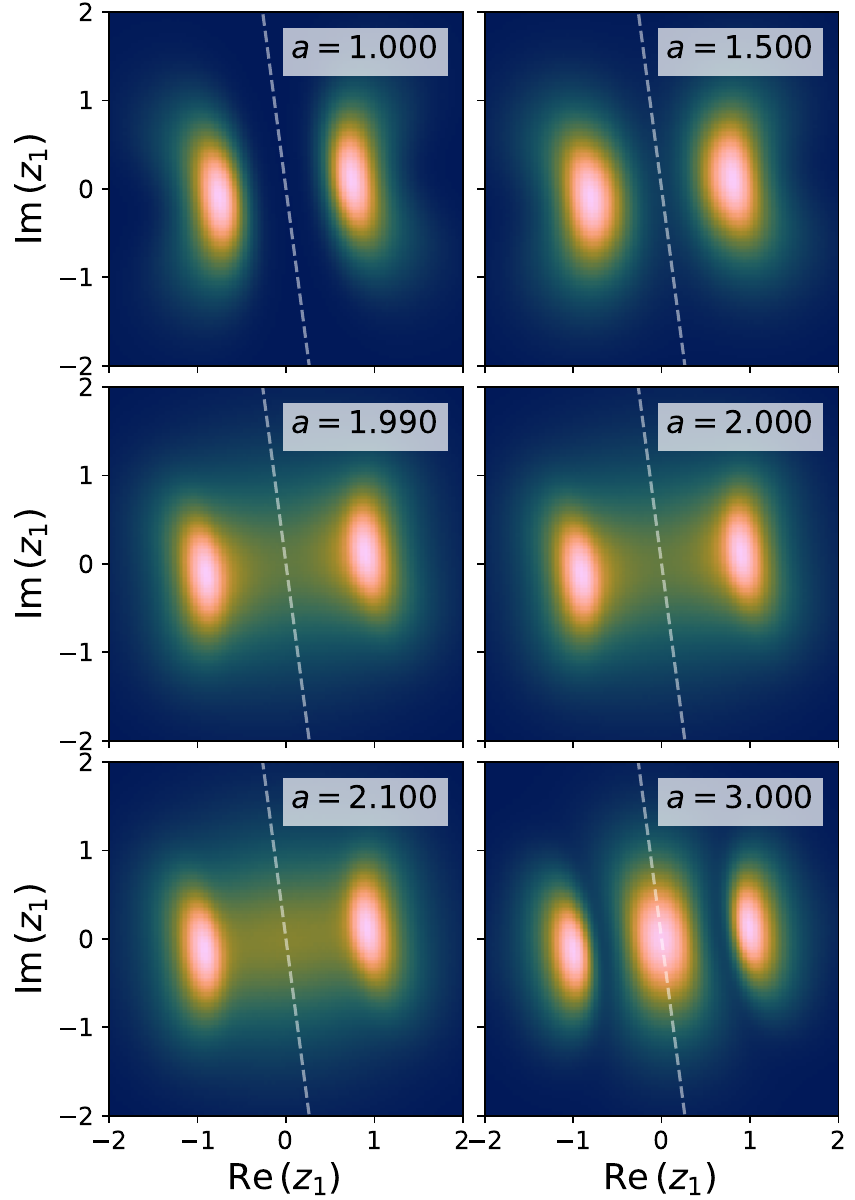}
    \caption{Histograms of $z_1$ obtained in complex Langevin simulations of the model  
             \eqref{eq:quartic_2d} with $\lambda=e^{5\ii\pi/6}$ and different values of the coupling 
             parameter $a$. We use equal kernels $K_1=K_2$, defined as in \eqref{eq:kernel_1d}, with 
             $m_i=22$, such that the respective distributions of $z_2$ are identical. The brighter regions 
             indicate a higher probability and the dashed lines lie in the direction of the argument of the 
             noise coefficients $\sqrt{K_i}$.}
\label{fig:histograms_2d_coupling_scan}
\end{figure}

For the case $m_1=m_2=22$, we show the histograms of $z_1$ and $z_2$ for different values of $a$ in 
\cref{fig:histograms_2d_coupling_scan}. One clearly observes how increasing the interaction strength induces 
a higher transition probability between the two high-probability regions. This even results in the curious 
observation that for the super-critical coupling $a=3.000$ there appear to be three instead of two high-
probability regions, all of which are parallel to $\sqrt{K_i}$. Note, however, that there is no significant 
change in the distributions in the vicinity of $a=2$.

At this point, we mention a peculiarity we have observed in our simulations: To begin with, we have 
performed simulations for values of $a$ equal to $0$, $0.5$, $1$, $1.5$, $1.99$, $2$, and $3$, supplemented by a
few smaller ensembles at $a=0.1$, $1.9$ and $2.1$,  and considered all possible combinations 
of $m_i\in\{10,22,34\}$ for each of them. Now, for all 
$a<2$, there were no major issues in any of the runs. As soon as we crossed the critical point $a=2$, 
however, we encountered problems that were not present before. In particular, for $a=2$, the simulations 
with $m_1=22\neq m_2$ or \emph{vice versa} slowed down drastically, preventing the algorithm from generating 
an ensemble of suitable size to perform averages over. This is likely caused by the adaptive step size 
algorithm producing smaller and smaller $\eps$ due to the drift terms blowing up. It also makes clear that 
one may not use arbitrary kernels for arbitrary theory parameters and still obtain viable results.
The failure of simulations for these particular combinations of $m_1$ and $m_2$ seems to be specific to 
$a=2$, however. For $a=2.1$ and $a=3$, on the other hand, we do not obtain results only when 
$m_1=10$, $m_2=34$ or \emph{vice versa}. 

\begin{figure}[t]
    \centering
    \includegraphics[width=1\linewidth]{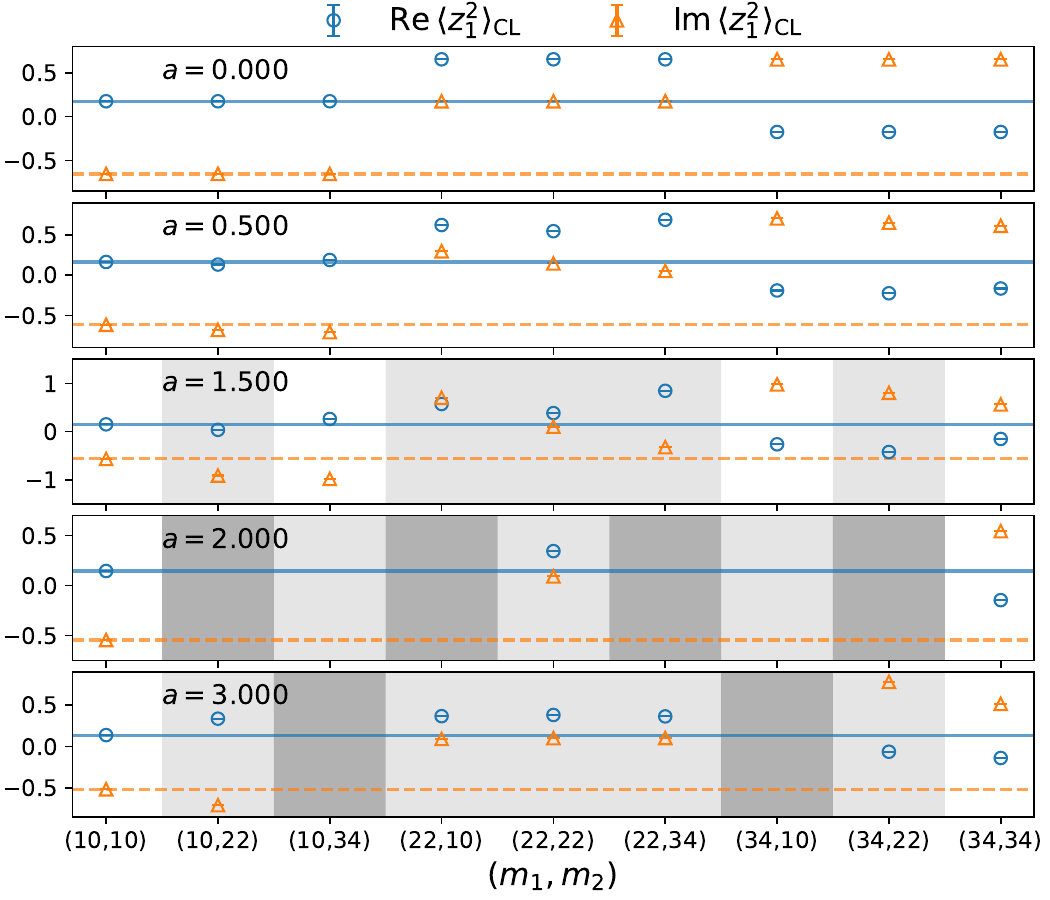}
    \caption{Real and imaginary parts of the observable $\langle z_1^2\rangle_{\mathrm{CL}}$ in the 
             theory \eqref{eq:quartic_2d} with $l=5$ in \eqref{eq:lambda} for different values of $a$. 
             The horizontal axis corresponds to different combinations of the integers $m_1$ and $m_2$, 
             parametrizing the kernel elements $K_1$ and $K_2$ as in \eqref{eq:kernel_1d}, respectively. For 
             combinations $(m_1,m_2)$ within the light shaded regions, there are nonzero boundary terms for 
             $\langle z_1^2\rangle_{\mathrm{CL}}$ and the dark shaded regions indicate those $(m_1,m_2)$ and $a$ 
             for which no simulation results could be obtained. The solid and 
             dashed horizontal lines indicate the real and imaginary parts of the corresponding exact 
             results, respectively, and the axis limits have been chosen such that all results without 
             boundary terms are fully visible.}
    \label{fig:observables_vs_m_2d_coupling_scan}
\end{figure}

To put the results discussed so far into context, we show in \cref{fig:observables_vs_m_2d_coupling_scan} 
the observable $\langle z_1^2\rangle_\CL$ for various values of $a$ and different combinations of $m_1$ and 
$m_2$. Those parameters for which we could not obtain results via simulations are marked by the dark shaded 
regions. Once again, we observe that while the presence of boundary terms indicates that 
$\langle z_1^{n_1}z_2^{n_2}\rangle_\CL\neq\langle z_1^{n_1}z_2^{n_2}\rangle_\mathrm{exact}$, the converse is 
not true, as some simulation results are seen to deviate from the exact ones despite boundary terms being 
consistent with zero. Other observables look qualitatively similar and are hence not shown here. 

In \cref{fig:observables_vs_m_2d_coupling_scan}, we trivially observe that there are no boundary 
terms in the free theory $a=0$ 
when the $m_i$ only take the values $10$, $22$, or $34$. This simply reflects the results of 
\cref{fig:observables_vs_m_1d}, since for $a=0$ we are simulating two independent 
one-dimensional theories of the form \eqref{eq:quartic_1d}. As one increases $a$, 
however, the interaction between $z_1$ and $z_2$ starts to become relevant, which has a 
nontrivial effect on the asymptotics of $S_a$ and, by extension, affects which contours are 
valid integration cycles. In turn, this has an influence on which trajectories, governed by 
the choice of $m_1$ and $m_2$, produce boundary terms. As seen in 
\cref{fig:observables_vs_m_2d_coupling_scan}, for $a\lesssim2$, only kernels with 
$m_i\in\{10,34\}$ cause boundary terms to vanish and we observe the onset of this behavior
around $a\approx1$. At and beyond the critical point $a=2$, we 
are then confronted with the situation that certain kernels do not produce usable results 
at all. Now, only the choices $(m_1,m_2)=(10,10)$ and $(34,34)$ ensure simulations free of
boundary terms. We expect these results not to change significantly when the kernel parameters 
$m_i$ are varied slightly, i.e., there should be plateaus in both $m_1$ and $m_2$ on 
which boundary terms vanish. In particular, we expect there to be a plateau around 
$(m_1,m_2)=(10,10)$, on which 
$\langle z_1^{n_1}z_2^{n_2}\rangle_\CL=\langle z_1^{n_1}z_2^{n_2}\rangle_{\mathrm{exact}}$.

Finally, we study how these choices of kernel affect the coefficients $a_i$ in 
\eqref{eq:salcedo_seiler}. We begin with the simpler case $a\geq2$. In this case, as we have
seen before, there are only $N_\gamma=2$ integration cycles, namely the real and imaginary
cycles $\gamma_1$ and $\gamma_2$, respectively. From \cref{fig:coefficients_vs_m_2d}, we
know that we may sample the former with choices of $(m_1,m_2)$ close to $(10,10)$ and the latter
with kernel parameters around $(34,34)$. On the other hand, for the combinations with
$m_1\neq m_2$ that we have studied, as well as for the choice $m_1=m_2=22$, we do not obtain
stable fit results, as in these cases we either find nonzero boundary terms or no results at all.

The case $a<2$ is less straightforward. First of all, since we now have to take into account 
$N_\gamma=9$ independent integration cycles, the set of five even observables 
\eqref{eq:observable_set_2d_even} is no longer sufficient to obtain a reasonable fit. Instead, 
we include all the observables we have measured, i.e., all $\langle z_1^{n_1}z_2^{n_2}\rangle_\CL$ 
with $n_1+n_2\leq4$. Next, we need to define a suitable basis set of integration cycles.
For $a=0$, in particular, we have seen above that one possible basis set consisting of nine 
cycles is given by the product $\gamma_i\times\gamma_j$, where $\gamma_i$ and $\gamma_j$ 
are one-dimensional cycles. In particular, we define
\begin{align}\label{eq:basis_cycles_2d}
    \begin{aligned}
        \gamma_1 := \gamma_1\times\gamma_1\;, \quad 
        \gamma_2 := \gamma_2\times\gamma_2\;, \quad 
        \gamma_3 := \gamma_1\times\gamma_2\;, \\
        \gamma_4 := \gamma_2\times\gamma_1\;, \quad 
        \gamma_5 := \gamma_1\times\gamma_3\;, \quad 
        \gamma_6 := \gamma_3\times\gamma_1\;, \\
        \gamma_7 := \gamma_2\times\gamma_3\;, \quad 
        \gamma_8 := \gamma_3\times\gamma_2\;, \quad 
        \gamma_9 := \gamma_3\times\gamma_3\;.
    \end{aligned}
\end{align}
Note that each of these cycles is really an equivalence class of integration surfaces in $\C^2$. 
Now, as is outlined in \cref{app:counting_cycles}, most of these equivalence classes shrink in
size as $a$ increases. However, one may nonetheless find certain representatives within them that 
constitute valid integration cycles for all $\vert a\vert<2$, such as the ones we use to compute
$\langle z_1^{n_1}z_2^{n_2}\rangle_{\gamma_i}$. Thus, in a slight abuse of notation, we shall 
consider the same basis \eqref{eq:basis_cycles_2d} for all $\vert a\vert<2$, keeping in mind that
one should restrict to certain subclasses of integration surfaces in practice. With this, we now 
compute the $a_i$ for values of the coupling $a<2$. 

We mention that in the cases where at least one of the distributions of $z_1$ or $z_2$ is confined
to a line, our fitting procedure cannot be applied due to the covariance matrix $\Sigma$ in 
\eqref{eq:least_squares} being singular. 
In these cases, however, the relevant integration cycles
are anyways easy to guess and the theorem \eqref{eq:salcedo_seiler} holds trivially. For $a=0$, this excludes all combinations $(m_1,m_2)$ except $(22,22)$
for the fit. For this kernel, we find that 
\begin{equation}
    a_1\approx\frac{\ii}{2}\;, \quad a_2\approx-\frac{\ii}{2}\;, \quad 
    a_3\approx\frac{1}{2}\;, \quad a_4=\frac{1}{2}
\end{equation}
within errors, whereas $a_{i>4}\approx0$, all in agreement with \eqref{eq:constraint}. This is
precisely the combination of coefficients one would have expected based on the results of 
\cref{fig:coefficients_vs_m_1d}. Also note that once more the real and imaginary parts of all 
coefficients are integer multiples of $1/2$.

\begin{figure}[t]
    \centering
    \includegraphics[width=1\linewidth]{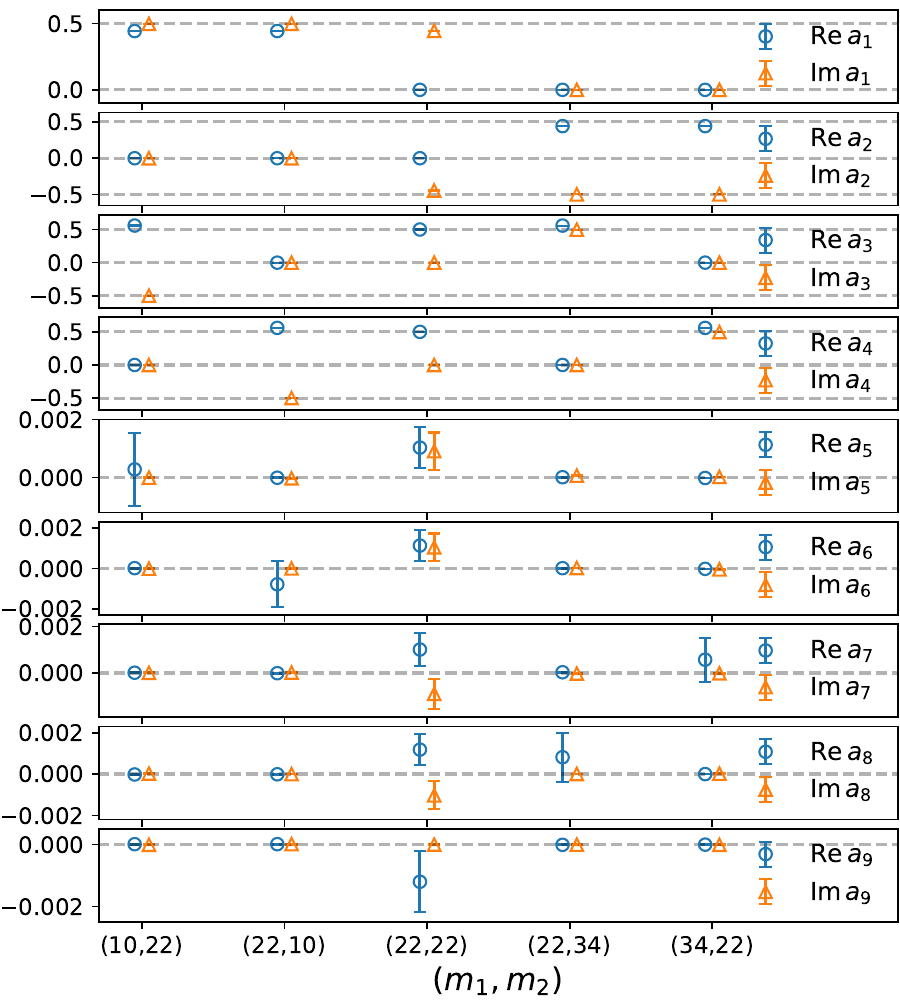}
    \caption{Real and imaginary parts of the coefficients $a_i$ in the model          
             \eqref{eq:quartic_2d} with $\lambda=e^{5\ii\pi/6}$ and $a=0.5$, obtained via the 
             least-squares fit procedure outlined in \cref{sec:measurement_coefficients} using 
             the observables $\langle z_1^{n_1}z_2^{n_2}\rangle$ with $n_1+n_2\leq4$. The horizontal 
             axis corresponds to different combinations of the integers $m_1$ and $m_2$, where
             we only show those combinations for which the distribution of neither $z_1$ nor $z_2$
             is confined to a line. The dashed horizontal lines are placed in steps of $1/2$ to 
             guide the eye and the data points have been displaced horizontally for better visibility. 
             Note that there are no boundary terms for any of the $(m_1,m_2)$ shown.}
    \label{fig:coefficients_vs_m_2d_unequal_kernels}
\end{figure}

Since in our one-dimensional simulations with a kernel of the form \eqref{eq:kernel_1d} we could
not sample $\gamma_3$, one is led to expect that with our choice of two-dimensional kernel, the 
linear combinations of cycles on the right-hand side of \eqref{eq:salcedo_seiler} are likely
restricted to $i\leq4$ for all $\vert a\vert<2$, as they are for $a=0$. What is more, the only
kernel for which we expect to sample all four of these integration cycles is precisely the one
given by $(m_1,m_2)=(22,22)$. For the other combinations of $m_1$ and $m_2$, at most two out of 
these four cycles should contribute. To substantiate this claim, we next investigate the case 
$a=0.5$. Owing to the smallness of the coupling, matters should still be qualitatively similar to
the free theory. Now, however, since the distributions of $z_1$ and $z_2$ are confined to lines 
only if both $m_i\in\{10,34\}$, c.f., \cref{fig:histograms_2d_unequal_kernels_a_1}, we may fit 
the $a_i$ for a larger set of kernels. 

We plot the coefficients for $a=0.5$ and different kernel parameters $(m_1,m_2)$ in 
\cref{fig:coefficients_vs_m_2d_unequal_kernels}. Indeed, we find four integration cycles
to be relevant when $(m_1,m_2)=(22,22)$ and two otherwise. For no choice of kernel do we find $\gamma_{i>4}$
to contribute significantly. We remark, however, that this might no longer be true when considering
more complicated kernels, such as full matrix kernels or kernels depending on the $z_i$ explicitly.
We are unable to perform an analogous analysis for the cases $a=1$ and $a=1.5$, as there are nonvanishing
boundary terms whenever at least one of the $m_i$ is equal to $22$, while the covariance matrices are singular in
all other cases. Finally, we note that, even though it is not easy to see in 
\cref{fig:coefficients_vs_m_2d_unequal_kernels} for nonzero couplings the real and imaginary parts of the 
coefficients are actually no longer restricted to being multiples of $1/2$, but start to deviate slightly.

%% file: conclusions.tex
We have investigated the problem of wrong convergence in complex Langevin simulations. In particular, we 
have studied the toy models \eqref{eq:quartic_1d} and \eqref{eq:quartic_2d} using various different kernels 
$K$ in the complex Langevin equation \eqref{eq:cle}. We found that, on the one hand, an appropriate choice 
of kernel can affect the distributions of the dynamical degrees of freedom in the complex plane in such a 
way that the measured monomial observables have vanishing boundary terms, a commonly employed correctness 
criterion in the complex Langevin approach. On the other hand, however, we have also shown that in some 
simulations without boundary terms the observables are still spoiled by unwanted integration cycles, in 
accordance with the theorem \eqref{eq:salcedo_seiler}.

We found that \eqref{eq:salcedo_seiler} holds in all our one- and two-dimensional simulations, indicating 
that there might be a 
straightforward extension of the proof of the theorem in \cite{SS19} to higher, and likely arbitrary, 
dimensions. We emphasize that contributions from unwanted integration cycles appear most prominently when a 
nontrivial kernel is used. In particular, our findings substantiate the claim that such spurious solutions 
to the complex Langevin equation should become relevant only in the presence of a nontrivial kernel or if 
the density $e^{-S}$ has singularities in the vicinity of the real axis \cite{Sal16}. 

As the second major result of this work, we have demonstrated how a kernel can be used -- in principle -- to 
control which integration cycles are sampled in a simulation. In particular, due to the deformation of 
distributions caused by nontrivial kernels, different cycles might be sampled for different arguments of 
$K$. Importantly, however, the introduction of a kernel may also make things worse in a sense. For instance, 
depending on the parameters of a theory, a kernel might force the simulation trajectories onto shapes which 
are not homotopic to any integration cycles of the theory. We have found evidence for this scenario to give 
rise to nonvanishing boundary terms. Even more importantly, the absence of boundary terms in the presence of 
a kernel does not guarantee correct results. While we have found large plateaus around the `perfect' choice 
of kernel on which the desired values of observables are reproduced, there are equally large plateaus with 
vanishing boundary terms for which one finds different results. 

At the moment, it is not clear to us to which extent our results can be generalized to more realistic 
theories, such as lattice models or even gauge theories. While one might expect the complexity (and the 
number) of integration cycles to increase the more degrees of freedom are involved, which would make an 
analysis like the one presented here impossible, we have also discussed a counterexample in this work. 
Namely, in the 
strong-coupling ($a\geq2$) regime of the two-dimensional model \eqref{eq:quartic_2d}, the analysis of 
integration cycles, in fact, becomes more straightforward than in the one-dimensional counterpart 
\eqref{eq:quartic_1d}. One can only speculate whether strongly coupled lattice theories with complicated 
interaction terms could actually feature a tractable number of integration cycles or one might even find 
$N_\gamma=1$. If the latter is not true, however, then any attempt of solving quantum field theories by means of the 
complex Langevin approach with a nontrivial kernel runs the risk of sampling unwanted integration cycles. 
We believe that this is something one should be aware of. In particular, it might be relevant for approaches 
in which one attempts to find suitable kernels by the means of machine learning \cite{LS23,ARS24}.

We plan to investigate this issue in more detail in future work. In particular, we have obtained 
preliminary results in a model similar to \eqref{eq:quartic_1d} but with multiple interaction terms, for 
which a $z$-independent kernel is no longer sufficient, but a $z$-dependent kernel producing correct results 
can nonetheless be found rather easily \cite{OSZ91}. Our results suggest that the findings presented here 
can be extended to that case rather straightforwardly. Notice, however, that in such simple theories it is 
possible to essentially guess what form an appropriate kernel should have. This, however, is far from 
straightforward in more general theories.

Another interesting research question is to study the connection between the sampling of unwanted 
integration cycles and the spectral properties of the underlying Fokker--Planck operator \cite{ALR23,SSS24}. 
In particular, in \cite{ALR23} correct results in a model similar to the ones studied in this work were 
associated with boundary terms being absent and the Fokker--Planck operator having only eigenvalues with a 
negative real part.

%% file: counting_cycles.tex
Contrary to the one-dimensional case discussed in \cref{sec:cycles_1d}, the counting of linearly independent 
integration cycles is nontrivial when two or more degrees of freedom are involved. This is due to 
the fact that there is no analog as simple as \eqref{eq:line_integrals} for the addition of two 
$d$-dimensional integrals. We thus devote this Appendix to developing a counting algorithm for 
integration cycles in arbitrary dimensions. However, we restrict ourselves to theories with 
regular actions $S$, which precludes the existence of closed-loop integration cycles. All theories 
with polynomial actions, for instance, are of this type. For a different approach of counting independent
integration cycles, see \cite{MHS24p}. 

Let us consider a theory with $d$ complex degrees of freedom $z_i$. In order to define integration 
cycles for such theories, we develop an intuitive language, which might not be the most elegant from a 
mathematical point of view, but which avoids the introduction of some rather advanced concepts of 
algebraic topology. The key point to note is that an integration cycle is a $d$-dimensional submanifold 
of the $2d$-dimensional space $\C^d$, with the important property that $e^{-S}$ vanishes on its 
boundaries. Since we consider only regular actions, those boundaries necessarily lie at complex 
infinity. More precisely, an integration cycle is actually an equivalence class of such manifolds, the
equivalence relation being that they share the same boundary conditions. For all practical purposes,
however, it is sufficient to consider only one representative out of each equivalence class and this 
shall be the goal in what follows. In particular, we aim at finding the simplest such representatives
possible.

To this end, we consider the subclass of integration manifolds within each equivalence class described by
a $d$-dimensional cycle $\gamma$ that can be written as $d$ independent contour integrals, i.e.,
\begin{equation}\label{eq:general_cycles}
    \int_{\gamma} d^dz = 
        \int_{a_i}^{b_1}dz_1\int_{a_2}^{b_2}dz_2\dots\int_{a_d}^{b_d}dz_d=:\int_{a}^{b}d^dz\;.
\end{equation}
Here, the $\int_{a_i}^{b_i}dz_i$ are, with a slight abuse of notation, defined to be integrals along 
arbitrary one-dimensional contours connecting the points $a_i$ and $b_i$ and we have introduced the 
right-hand side of \eqref{eq:general_cycles} as a shorthand notation, with $a_i$ being the components 
of $a$ and similar for $b$. Needless to say, in order for $\gamma$ to constitute a valid integration 
cycle in a theory with an action $S$, the integral \eqref{eq:general_cycles} with the integrand $e^{-S}$ 
must exist. A necessary condition for this is that the points $a$ and $b$ both lie in regions at complex 
infinity in which $e^{-S}$ vanishes. To be more concrete, let us introduce polar coordinates, 
$z_i=r_ie^{\ii\theta_i}$, as in \cref{sec:cycles_1d} and define $r^2 := \sum_{i=1}^dr_i^2$. We can 
then again identify disjoint regions $G_i$ in the space spanned by the $\theta_i$ (with periodic 
boundary conditions), i.e., the torus $\Theta_d:=[0,2\pi)^d$, in which $e^{-S}\to0$ as $r\to\infty$. 
Let us define the tuple of complex arguments of $z$ as $\arg\,z:=(\arg\,z_1,\dots,\arg\,z_d)$. Then, 
in light of \eqref{eq:general_cycles}, we require that $\arg\,a\in G_i$ while $\arg\,b\in G_j$ 
with $i\neq j$, and that $\sum_{i=1}^d\vert a_i\vert^2=\infty$ and analogously for $b$. 

\begin{figure*}[t]
    \centering
    \includegraphics[width=0.9\linewidth]{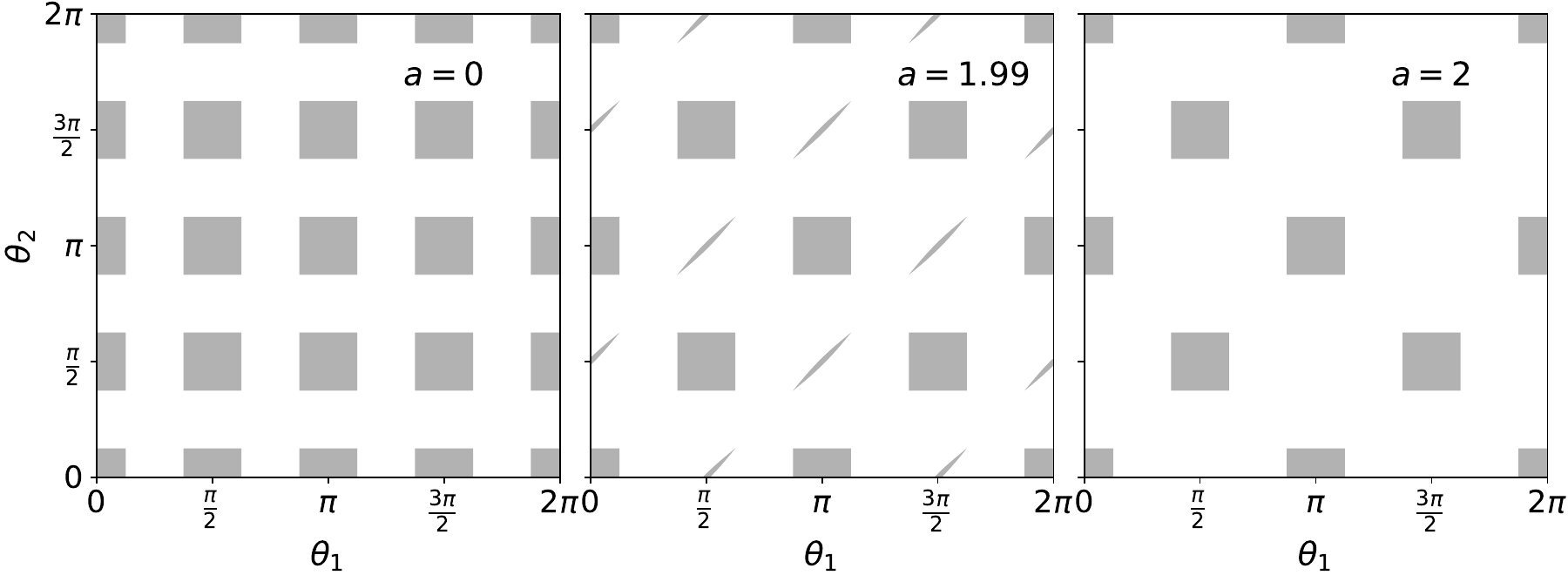}
    \caption{`Sign' of $f_\phi(\theta_1,\theta_2)$ in the $(\theta_1,\theta_2)$ plane for different 
             $a$ in \eqref{eq:quartic_2d}. In the shaded regions, $f_\phi(\theta_1,\theta_2)>0$ for 
             all $\phi\in[0,\frac{\pi}{2}]$ such that $e^{-S(z_1,z_2)}\to0$ in the limit $r\to\infty$. 
             For $\vert a\vert >2$, analogous plots are identical to the right panel.}
    \label{fig:zeros_2d}
\end{figure*}

Although this is a necessary condition for \eqref{eq:general_cycles} to exist, it is not a sufficient one. 
To see this, consider exchanging any of the $a_i$ with the corresponding $b_i$, resulting in a
sign flip of the integral \eqref{eq:general_cycles}, i.e., 
\begin{equation}
    \int_{a_1}^{b_1}dz_1\dots\int_{b_i}^{a_i}dz_i\dots\int_{a_d}^{b_d}dz_d=-\int_a^bd^dz\;.
\end{equation}
Obviously, if the integral on the right should exist, then so must the one on the left. This, however, 
has to be true for any number of exchanges between the upper and lower limits in the integral, as it 
changes the result at most by a sign. This imposes severe restrictions on the possible $a$ and $b$ that 
give rise to valid integration cycles. More concretely, let us define the set of points whose $i$-th 
coordinates are either $a_i$ or $b_i$,
\begin{equation}\label{eq:tuple_set}
    X := \left\{(x_1,x_2,\dots,x_d)\,\big\vert\, x_i\in\left\{a_i,b_i\right\}\right\}\;.
\end{equation}
Then, what is required for $\int_a^bd^dz$ to define a valid representative of a cycle $\gamma$ is 
that the tuples $\arg\,x$ for the different $x\in X$ all lie in different regions $G_i$.

In the following, we shall develop a graphical interpretation of this picture that is based on rectangular
cuboids in the space $\Theta_d$, namely those cuboids whose vertices are given by $\arg\,x$ for $x\in X$ 
and whose sides are thus parallel to the coordinate axes. Indeed, since the regions $G_i$ are disjoint 
subsets of $\Theta_d$, a sufficient requirement for the integral $\int_a^b d^dz$ to represent an integration 
cycle is that the vertices of the cuboid defined by $X$ are each contained in a different $G_i$. Note that 
$\arg\,a$ and $\arg\,b$ are two diagonally opposite vertices of this cuboid. Thus, any other two points 
$a'$ and $b'$ whose arguments lie on diagonally opposite vertices of the same cuboid define an integration 
manifold within the same equivalence class of cycles. Moreover, the parallel shift of vertices within the 
$G_i$ corresponds to a homotopy transformation and thus does not give rise to a different integration cycle. 
In other words, a single cuboid describes multiple (but equivalent) possible integration manifolds. 
Furthermore, if a cuboid is degenerate, i.e., if two or more of its vertices happen to lie within the same 
$G_i$ (which is the case if one or more $a_i=b_i$), the resulting integral is still well-defined but vanishes 
exactly and is thus not relevant for our purposes.

Let us demonstrate the idea, considering the class of two-dimensional theories defined in \eqref{eq:quartic_2d} 
with $\lambda=4$ as an example. After introducing polar coordinates and defining the angle 
$\phi := \arctan\left(\frac{r_2}{r_1}\right)\in[0,\frac{\pi}{2}]$, we can write the real part of $S(z)$ as
\begin{equation}
    \re\,S(z) = r^4f_{\phi}(\theta_1,\theta_2)\;,
\end{equation}
with
\begin{align}
    \begin{aligned}
        f_\phi(\theta_1,\theta_2) = \cos^4\left(\phi\right)\bigg[&\cos(4\theta_1) + 
		  					           \tan^4\left(\phi\right)\cos(4\theta_2) \\ 
                                    &+a\tan^2\left(\phi\right)\cos\left(2(\theta_1+\theta_2)\right)\bigg]\;.
    \end{aligned}
\end{align}
The problem of determining the regions $G_i$ where $e^{-S}\to0$ (i.e., $\re\,S\to\infty$) as $r\to\infty$ 
thus reduces to the problem of finding the regions in $(\theta_1,\theta_2)$-space in which 
$f_\phi(\theta_1,\theta_2)>0$ for all $\phi\in[0,\frac{\pi}{2}]$. We show these regions for three 
different values of the coupling parameter $a$ in \cref{fig:zeros_2d}. As can be seen, the number $N_G$ of 
regions $G_i$ changes as $a$ is varied. Indeed, we find $N_G=16$ if $-2<a<2$ and $N_G=8$ otherwise, taking 
into account the periodicity in both $\theta_1$ and $\theta_2$. As is explained below, 
this fact results in a jump in the number of linearly independent integration cycles at $a=\pm2$. The 
construction of the $G_i$ in more general theories is analogous. 

We remark at this point that the fact that $f_\phi(\theta_1,\theta_2)$ must be positive for \emph{all} 
values of $\phi$ is important. Indeed, while one might naively expect to be able to consider the interval 
$[0,\frac{\pi}{2}]$ only up to a set of measure zero, this is not true in general. In fact, without going 
into the details, we have found in the case $a=2$ that for $\phi=\frac{\pi}{4}$, corresponding to the line 
$r_1=r_2$, a non-integrable singularity arises in the integral over certain cycles. Thus, simply omitting 
this point would lead to the wrong conclusion that there are $9$ independent cycles for $a=2$, when there 
are really only two, namely those unaffected by the aforementioned singularity. Following the above argument, 
the allowed integration cycles of \eqref{eq:quartic_2d} can now be represented as rectangles whose vertices 
all lie in different $G_i$. For $a=2$, as seen in the right panel of \cref{fig:zeros_2d}, there are only two 
such rectangles, from which we conclude that the theory \eqref{eq:quartic_2d_o2} has at most two independent 
integration cycles.

\begin{figure*}[t]
    \centering
    \includegraphics[width=0.9\linewidth]{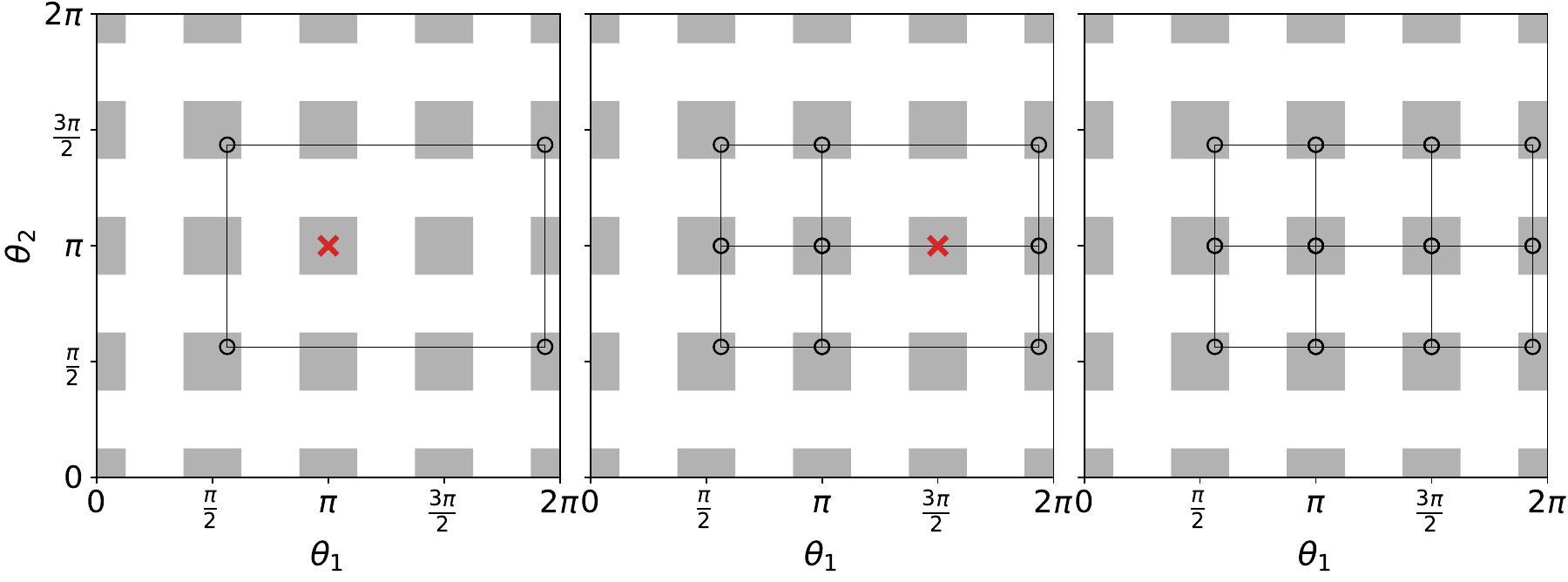}
    \caption{As in the left panel of \cref{fig:zeros_2d} but in addition we show how our algorithm can be 
             used to decompose rectangles into smaller ones: Starting from the rectangle in the left panel, 
             we choose a $c$ in its interior (red cross) to obtain the four smaller rectangles shown in the 
             center panel. The smaller two of these rectangles cannot be decomposed anymore, while for the 
             other two we choose a $c$ that lies on the side they share (red cross in the center panel). This 
             gives rise to four additional rectangles, as seen in the right panel. The resulting six 
             rectangles cannot be decomposed anymore, i.e., they are atomic, such that the algorithm has 
             converged.}
    \label{fig:algorithm}
\end{figure*}

The final ingredient that we need in order to reduce the set of all possible integration cycles to a set of 
linearly independent ones is a generalization of the addition rule \eqref{eq:line_integrals} for higher 
dimensions. With an arbitrary (possibly infinite) point $c$, such a generalization can be written as
\begin{equation}\label{eq:general_line_integals}
    \int_{a}^{b}d^dz = \sum_{x\in X}\sign(x)\int_{x}^{c}d^dz\;,
\end{equation}
where $X$ is the set defined in \eqref{eq:tuple_set} and we have introduced the sign function on $X$ as
\begin{equation}
    \sign(x) := \prod_{i=1}^d s_i\;, \quad s_i = \begin{cases}+1 \ \mathrm{if}\ x_i=a_i\\
                       -1 \ \mathrm{if}\ x_i=b_i
          \end{cases}\;.
\end{equation}
With this, the increase in complexity when going beyond $d=1$ (where the number of independent integration 
cycles for an $n$-th-order-polynomial action is simply given by $n-1$) can easily be appreciated. Indeed, 
while the right-hand side of \eqref{eq:general_line_integals} consists of only two terms for $d=1$ (c.f.  
\eqref{eq:line_integrals}), it involves $2^d$ terms in general.

The graphical interpretation of \eqref{eq:general_line_integals} is that one may decompose any integration
cycle, i.e., any cuboid, into a sum (with appropriate signs) of all cuboids that share one vertex with the 
original cuboid and have the point $c$ as its diagonally opposite vertex. This observation is at the heart 
of our counting algorithm. 

Indeed, the underlying idea is to start with any cuboid that represents a valid integration cycle and use 
\eqref{eq:general_line_integals} to decompose it into smaller and smaller cuboids. Here, by the `size' of 
a cuboid we refer to its volume in $\Theta_d$. For instance, one could consider the smallest volume of all 
cuboids within the same equivalence class. Such a reduction can be accomplished by choosing the point $c$ 
in \eqref{eq:general_line_integals} to lie within (or on) the original cuboid, giving rise to $2^d$ smaller 
cuboids, each of which one again tries do decompose further until convergence. Notice that this procedure
might give rise to degenerate cuboids, which correspond to vanishing integrals and can thus simply be 
discarded. At the end of such an iterative procedure, one is left with a set of small cuboids that cannot 
be decomposed anymore and that we thus henceforth refer to as atomic. In other words, the original 
cuboid can be represented as a sum (with appropriate signs) of these atomic cuboids and such a decomposition 
in terms of atomics is possible for any given cuboid. Thus, the atomic cuboids constitute a basis for the 
cuboids, i.e., for integration cycles. Importantly, this algorithm is guaranteed to converge as the size 
of the involved cuboids decreases with every iteration step and there is a minimum size of cuboids for a 
given theory. 

An example of how this algorithm works within the two-dimensional theory \eqref{eq:quartic_2d} with $a=0$ 
is shown in \cref{fig:algorithm}, where the rectangle on the left plot is decomposed into atomic rectangles 
after two iteration steps. It is not hard to see that in total there are nine such atomic rectangles in 
this theory, from which we conclude that there are nine linearly independent integration cycles. The same 
is true for all $\vert a\vert<2$, as can be seen, for instance, in the center plot of \cref{fig:zeros_2d}, 
where we consider $a=1.99$. For $a\geq2$, however, there are only two atomic rectangles. This is because 
eight of the regions $G_i$ decrease in size as $\vert a\vert$ increases from $0$ to $2$, causing them to 
disappear completely at $a=2$, leaving only two independent cycles. Notice that this shrinking of the $G_i$ 
also implies that not all possible integration manifolds that are valid cycles for $a=0$ are also valid 
for $0<a<2$. However, there is at least one representative out of each equivalence class that defines an 
allowed cycle for all $\vert a\vert\leq2$. For instance, such a representative is given by the cuboid 
that connects the centers of four different $G_i$, see \cref{fig:zeros_2d}.

We note that this algorithm is capable of finding one particular set of linearly independent integration 
cycles. In practice, however, this basis is not always the most convenient to use, such that one may resort 
to using different sets instead, as is done, for instance, in \cref{sec:results_2d}. Nonetheless, it can 
be a helpful tool in determining the number of independent integration cycles to use, for instance, in a 
computer program and the generalization to higher dimensions $d$ is straightforward. We also note that if 
one knows a set of possible integration cycles $\gamma_j$ of a given theory and can furthermore compute 
expectation values of a sufficiently large set of observables $\obs_i$ along these cycles, then the rank 
of the matrix $M$, defined by $M_{ij}:=\langle\obs_i\rangle_{\gamma_j}$, also gives (a lower bound on) the 
number of independent integration cycles. For the cases studied in this work, we have verified that this 
method and the above algorithm give consistent results. 

%% file: chi_squared.tex
In order to compute estimators for expectation values and uncertainties of the coefficients 
$a_i$ in \eqref{eq:salcedo_seiler}, we perform the least-squares fit analysis described in 
\cref{sec:measurement_coefficients}. For the interpretation of results, however, assessing the 
goodness of such a fit is of fundamental importance. In this appendix, we shall be concerned 
with this question. In particular, our aim is to find a measure for the extent to which 
\eqref{eq:salcedo_seiler} is fulfilled. Thereby, we compare the observables 
$\langle\obs\rangle_\CL$ measured in our complex Langevin simulations to the model 
predictions $\sum_{i=1}^{N_\gamma}a_i\langle\obs\rangle_{\gamma_i}$, where the $a_i$ are the 
coefficients resulting from the fit procedure and $\langle\obs\rangle_{\gamma_i}$ are the 
expectation values of observables $\obs$ along the integration cycles $\gamma_i$, which, in 
the models we investigate in this work, we can compute via numerical integration. As in 
\cref{sec:measurement_coefficients}, however, we consider only real observables, such that 
what we are really concerned with is the deviation between the vectors $\boldsymbol{y}$ and 
$X\boldsymbol{\beta}$, where $\boldsymbol{y}$, $X$ and $\boldsymbol{\beta}$ are all 
defined in \cref{sec:measurement_coefficients}.

To begin with, perhaps the most common goodness-of-fit test is the reduced $\chi^2$ statistic. 
For this, one considers the vector of residuals, i.e., $\boldsymbol{y}-X\boldsymbol{\beta}$, 
and computes the sum of their squares weighted by the (suitably normalized) covariance matrix 
of $\boldsymbol{y}$, $\Sigma$, c.f., \eqref{eq:chi_squared}:
\begin{equation}
   \chi^2 := 
    (\boldsymbol{y}-X\boldsymbol{\beta})^T\Sigma^{-1}(\boldsymbol{y}-X\boldsymbol{\beta})\;.
\end{equation}
The value of $\chi^2$ is then compared to the effective number of (in this case: real) degrees 
of freedom, $N_{\mathrm{dof}}=2N_\obs-2N_\gamma$.
In particular, one concludes that the fit is good (bad) if $\chi^2\approx N_{\mathrm{dof}}$
($\chi^2\gg N_{\mathrm{dof}}$), while it is considered an overfit if 
$\chi^2<N_{\mathrm{dof}}$. The issue with this procedure, however, is that it relies on the 
residuals being Gaussian distributed. As we shall see below, this condition is not always met 
for our data, preventing us from using the standard 
reduced $\chi^2$ statistic as a goodness-of-fit test in a straightforward way.

Instead, we follow \cite{ASM10u} by studying the distribution of the weighted residuals directly.
Under the assumption that the model is correct and the data are uncorrelated and follow a 
Gaussian distribution with known variances $\sigma_i^2$, then the residuals, weighted by 
$1/\sigma_i$, should follow a standard normal distribution. Since the data in our simulations are
not uncorrelated, however, we first decorrelate the measurements by performing a so-called whitening
transformation. There are infinitely many ways to do this in general \cite{KLS18}. Here, we
consider one particular method, namely the \emph{ZCA-cor} whitening transformation,
retaining the maximal amount of component-wise correlation between the original and the whitened
data \cite{KLS18}. In this approach, the whitened residuals are defined as
\begin{equation}\label{eq:whitened_residuals}
    \boldsymbol{x} = W^{\mathrm{ZCA-cor}}(\boldsymbol{y}-X\boldsymbol{\beta})\;, 
\end{equation}    
with the whitening matrix
\begin{equation}
    W^{\mathrm{ZCA-cor}} = P^{-1/2}V^{-1/2}\;.
\end{equation}
Here, the matrices $P$ and $V$ are defined such that $\Sigma=V^{1/2}PV^{1/2}$, where 
$\Sigma$ again denotes the covariance matrix of the measurements $\boldsymbol{y}$, 
but now it is computed as an average over simulation runs and not additionally normalized.
In other words, $P$ corresponds to the correlation matrix, while $V$ is the diagonal 
variance matrix. Similarly, $\boldsymbol{\beta}$ in \eqref{eq:whitened_residuals} denotes the 
average over runs as well. Note that the whitening matrix is computed from the full data set once,
while we compute one vector $\boldsymbol{x}$ from $\boldsymbol{y}$ for every simulation run, leaving
us with a total $2N_\obs\Nruns$ data points $x_i$. Indeed, we find each $\boldsymbol{x}$ to be perfectly 
uncorrelated, such that we may study the distribution of the $x_i$ as a test for the goodness of our fitting 
procedure.

\begin{figure}[t]
    \centering
    \includegraphics[width=1\linewidth]{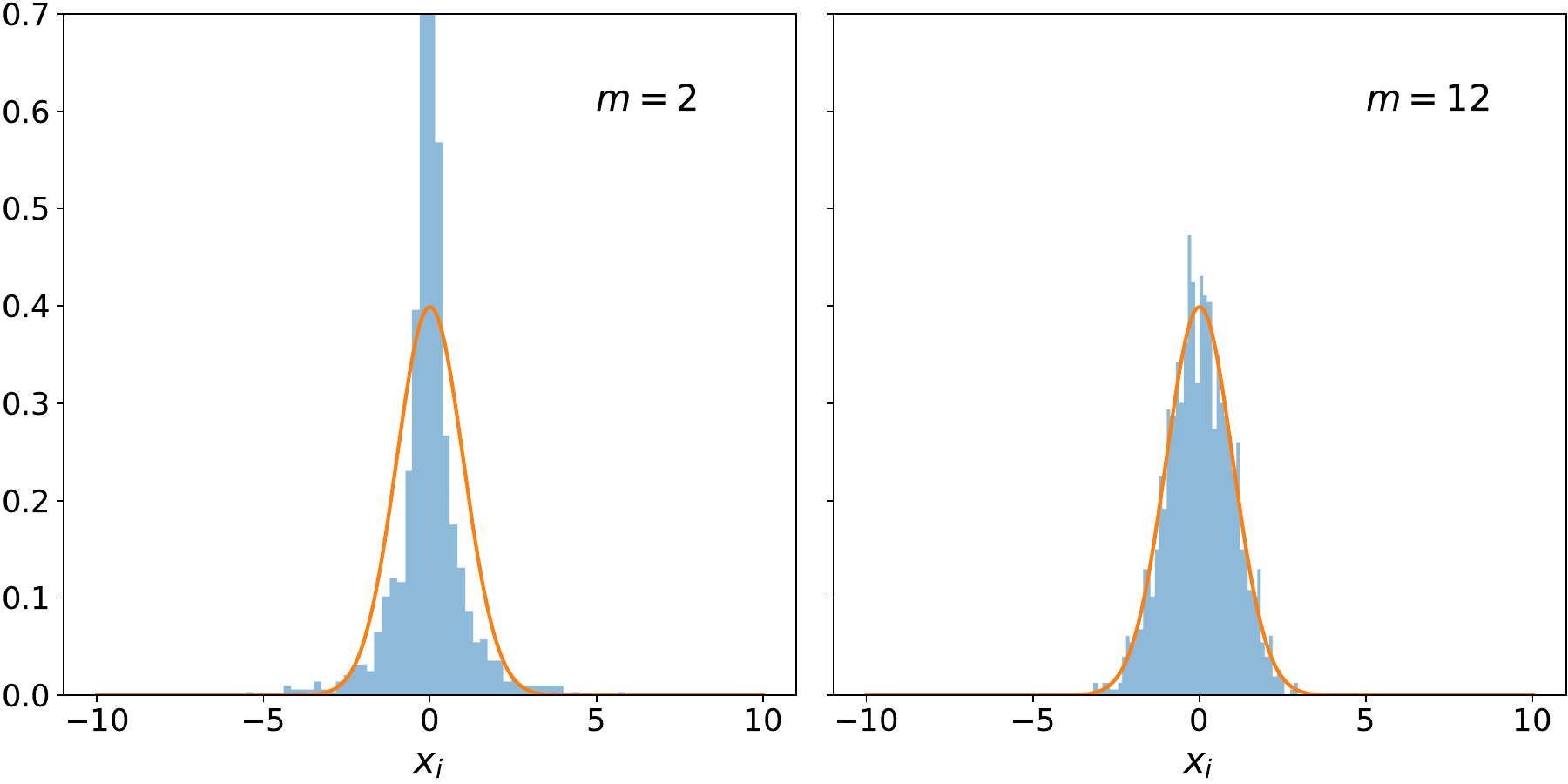}
    \caption{Distribution of the whitened residuals $x_i$ in \eqref{eq:whitened_residuals} obtained
             from complex Langevin simulations of the model \eqref{eq:quartic_1d} with $\lambda=e^{5\ii\pi/6}$
             and a kernel of the form \eqref{eq:kernel_1d} for two different values of the 
             integer $m$. The histograms have been normalized such that they sum to $1$ and a
             normal distribution with zero mean and unit variance is shown for comparison.}
    \label{fig:residual_distributions}
\end{figure}

Two exemplary such distributions that we obtain in the theory \eqref{eq:quartic_1d} with 
$\lambda=e^{5\ii\pi/6}$ for a kernel of the form \eqref{eq:kernel_1d} with $m=2$ and $m=12$, 
respectively, are shown in \cref{fig:residual_distributions}.
Indeed, we see that the distribution closely resembles a Gaussian with zero mean and unit 
variance for $m=12$, where
we know boundary terms to be consistent with zero (c.f. \cref{fig:observables_vs_m_1d}), 
whereas there is a substantial deviation for $m=2$, where boundary terms are nonzero.

Lastly, in order to quantify the deviation of the produced distribution from a standard normal distribution,
we employ the Shapiro--Wilk test \cite{SW65}, from which we may compute the associated $p$-value. 
This quantity is the probability of measuring the observed distribution under
the assumption that the null hypothesis, i.e., that the whitened residuals are drawn from a
standard Gaussian distribution, is correct. A small $p$-value thus indicates that the observed
outcome is likely incompatible with the residuals being Gaussian distributed.
In turn, this means from that large (small) $p$-values within the Shapiro--Wilk test, we may conclude that 
the fit for the coefficients $a_i$ is good (bad). As a threshold, we use the commonly employed value $p=0.05$.

\begin{figure}[t]
    \centering
    \includegraphics[width=1\linewidth]{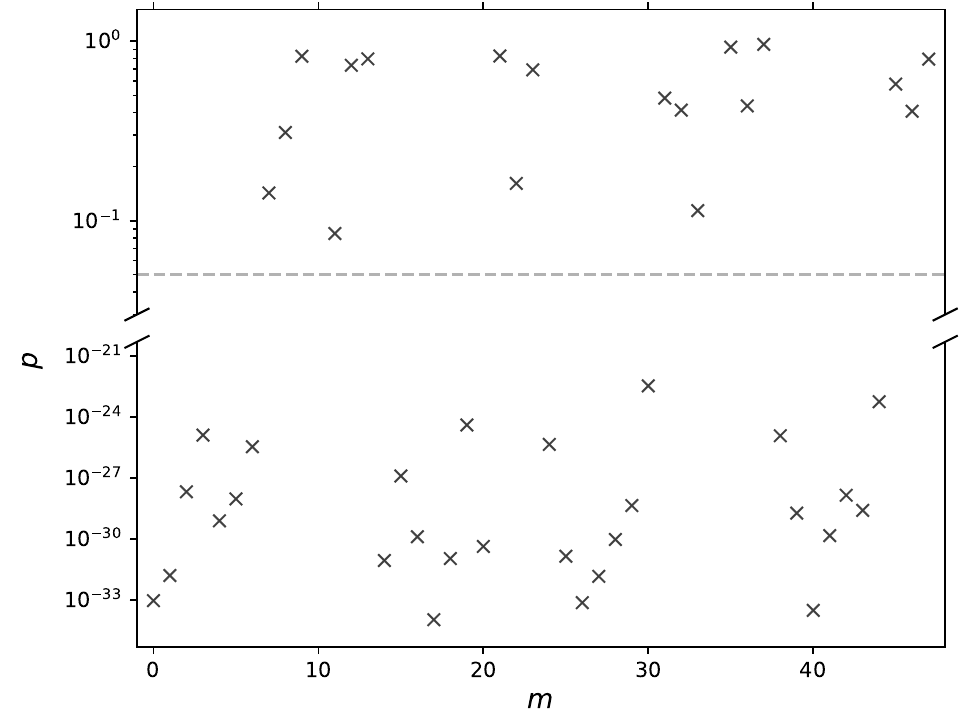}
    \caption{$p$-value of a Shapiro--Wilk normality test applied to the whitened residuals obtained
             from complex Langevin simulations of the model \eqref{eq:quartic_1d} with $\lambda=e^{5\ii\pi/6}$.
             The horizontal axis is the kernel parameter $m$ in \eqref{eq:kernel_1d} and the threshold 
             value $0.05$ is shown as the horizontal dashed line. Notice the interrupted vertical axis and
             the logarithmic scale.}
    \label{fig:p_value_vs_m_1d}
\end{figure}

For the model \eqref{eq:quartic_1d} with $\lambda=e^{5\ii\pi/6}$, we show the $p$-value as a function of the 
kernel parameter $m$ in \cref{fig:p_value_vs_m_1d}. Comparing this result to \cref{fig:observables_vs_m_1d}, 
we observe a clear correlation between the presence of boundary terms and $p$-values (many orders of magnitude) 
below $0.05$. 
This is, in fact,  true in all of our simulations, both in one and two dimensions. We have also 
experimented with different whitening transformations in \eqref{eq:whitened_residuals} and obtained 
consistent results. In conclusion, we find the approach presented here to provide a viable criterion for 
assessing the goodness of our least-squares fits in contrast to the reduced $\chi^2$ statistic.

